\documentclass[particles,article,submit,pdftex,moreauthors]{Definitions/mdpi} 
\firstpage{1} 
\pubvolume{1}
\issuenum{1}
\articlenumber{0}
\pubyear{2025}
\copyrightyear{2025}
\externaleditor{Matteo Duranti and Valerio Vagelli} % More than 1 editor, please add `` and '' before the last editor name
\datereceived{ } 
\daterevised{ } % Comment out if no revised date
\dateaccepted{ } 
\datepublished{ } 
\hreflink{https://doi.org/} % If needed use \linebreak
\graphicspath{{./Pictures/}}

\Title{Spectral performance of single-channel plastic and GAGG scintillator bars of the CUbesat Solar Polarimeter (CUSP)}

\TitleCitation{Spectral performance of single-channel plastic and GAGG scintillator bars of the CUbesat Solar Polarimeter (CUSP)}

\Author{
    Nicolas De Angelis$^{1,\ast}$\orcidA{},
    Abhay Kumar$^{1}$,
    Sergio Fabiani$^{1}$\orcidB{},
    Ettore Del Monte$^{1}$\orcidC{},
    Enrico Costa$^{1}$\orcidD{},
    Giovanni Lombardi$^{1,2}$\orcidE{},
    Alda Rubini$^{1}$,
    Paolo Soffitta$^{1}$\orcidF{},
    Andrea Alimenti$^{1,3}$\orcidG{},
    Riccardo Campana$^{4,5}$\orcidH{},
    Mauro Centrone$^{6}$\orcidI{},
    Giovanni De Cesare$^{4}$\orcidJ{},
    Sergio Di Cosimo$^{1}$,
    Giuseppe Di Persio$^{1}$,
    Alessandro Lacerenza$^{1}$,
    Pasqualino Loffredo$^{1}$,
    Gabriele Minervini$^{7}$,
    Fabio Muleri$^{1}$\orcidK{},
    Paolo Romano$^{8}$\orcidL{},
    Emanuele Scalise$^{1}$,
    Enrico Silva$^{1,3}$,
    Davide Albanesi$^{9}$,
    Ilaria Baffo$^{10}$,
    Daniele Brienza$^{11}$,
    Valerio Campomaggiore$^{9}$,
    Giovanni Cucinella$^{12}$,
    Andrea Curatolo$^{13}$,
    Giulia de Iulis$^{9}$,
    Andrea Del Re$^{9}$,
    Vito Di Bari$^{12}$,
    Simone Di Filippo$^{12}$,
    Immacolata Donnarumma$^{11}$\orcidM{},
    Pierluigi Fanelli$^{10}$,
    Nicolas Gagliardi$^{14}$,
    Paolo Leonetti$^{9}$,
    Matteo Mergè$^{11}$,
    Dario Modenini$^{13,14}$\orcidN{},
    Andrea Negri$^{12}$,
    Daniele Pecorella$^{13}$,
    Massimo Perelli$^{12}$,
    Alice Ponti$^{14}$,
    Francesca Sbop$^{9}$,
    Paolo Tortora$^{13,14}$,
    Alessandro Turchi$^{11}$,
    Valerio Vagelli$^{11}$,
    Emanuele Zaccagnino$^{11}$,
    Alessandro Zambardi$^{9}$ and
    Costantino Zazza$^{15}$
}
\AuthorNames{
    Nicolas De Angelis, Abhay Kumar, Sergio Fabiani, Ettore Del Monte, Enrico Costa, Giovanni Lombardi, Alda Rubini, Paolo Soffitta, Andrea Alimenti, Riccardo Campana, Mauro Centrone, Giovanni De Cesare, Sergio Di Cosimo, Giuseppe Di Persio, Alessandro Lacerenza, Pasqualino Loffredo, Gabriele Minervini, Fabio Muleri, Paolo Romano, Emanuele Scalise, Enrico Silva, Davide Albanesi, Ilaria Baffo, Daniele Brienza, Valerio Campamaggiore, Giovanni Cucinella, Andrea Curatolo, Giulia de Iulis, Andrea Del Re, Vito Di Bari, Simone Di Filippo, Immacolata Donnarumma, Pierluigi Fanelli, Nicolas Gagliardi, Paolo Leonetti, Matteo Mergè, Dario Modenini, Andrea Negri, Daniele Pecorella, Massimo Perelli, Alice Ponti, Francesca Sbop, Paolo Tortora, Alessandro Turchi, Valerio Vagelli, Emanuele Zaccagnino, Alessandro Zambardi, Costantino Zazza
}
\isAPAStyle{%
       \AuthorCitation{
        De Angelis, N., Kumar, A., Fabiani, S., Del Monte, E., Costa, E., Lombardi, G., Rubini, A., Soffitta, P., Alimenti, A., Campana, R., Centrone, M., De Cesare, G., Di Cosimo, S., Di Persio, G., Lacerenza, A., Loffredo, P., Minervini, G., Muleri, F., Romano, P., Scalise, E., Silva, E., Albanesi, D., Baffo, I., Brienza, D., Campamaggiore, V., Cucinella, G., Curatolo, A., de Iulis, G., Del Re, A., Di Bari, V., Di Filippo, S., Donnarumma, I., Fanelli, P., Gagliardi, N., Leonetti, P., Mergè, M., Modenini, D., Negri, A., Pecorella, D., Perelli, M., Ponti, A., Sbop, F., Tortora, P., Turchi, A., Vagelli, V., Zaccagnino, E., Zambardi, A., \& Zazza, C.
        }
         }{%
        \isChicagoStyle{%
        \AuthorCitation{Nicolas De Angelis, Abhay Kumar, Sergio Fabiani, Ettore Del Monte, Enrico Costa, Giovanni Lombardi, Alda Rubini, Paolo Soffitta, Andrea Alimenti, Riccardo Campana, Mauro Centrone, Giovanni De Cesare, Sergio Di Cosimo, Giuseppe Di Persio, Alessandro Lacerenza, Pasqualino Loffredo, Gabriele Minervini, Fabio Muleri, Paolo Romano, Emanuele Scalise, Enrico Silva, Davide Albanesi, Ilaria Baffo, Daniele Brienza, Valerio Campamaggiore, Giovanni Cucinella, Andrea Curatolo, Giulia de Iulis, Andrea Del Re, Vito Di Bari, Simone Di Filippo, Immacolata Donnarumma, Pierluigi Fanelli, Nicolas Gagliardi, Paolo Leonetti, Matteo Mergè, Dario Modenini, Andrea Negri, Daniele Pecorella, Massimo Perelli, Alice Ponti, Francesca Sbop, Paolo Tortora, Alessandro Turchi, Valerio Vagelli, Emanuele Zaccagnino, Alessandro Zambardi, and Costantino Zazza.
}
        }{
        \AuthorCitation{
        De Angelis, N.; Kumar, A.; Fabiani, S.; Del Monte, E.; Costa, E.; Lombardi, G.; Rubini, A.; Soffitta, P.; Alimenti, A.; Campana, R.; Centrone, M.; De Cesare, G.; Di Cosimo, S.; Di Persio, G.; Lacerenza, A.; Loffredo, P.; Minervini, G.; Muleri, F.; Romano, P.; Scalise, E.; Silva, E.; Albanesi, D.; Baffo, I.; Brienza, D.; Campamaggiore, V.; Cucinella, G.; Curatolo, A.; de Iulis, G.; Del Re, A.; Di Bari, V.; Di Filippo, S.; Donnarumma, I.; Fanelli, P.; Gagliardi, N.; Leonetti, P.; Mergè, M.; Modenini, D.; Negri, A.; Pecorella, D.; Perelli, M.; Ponti, A.; Sbop, F.; Tortora, P.; Turchi, A.; Vagelli, V.; Zaccagnino, E.; Zambardi, A.; Zazza, C.
        }
        }
}

\address{%
$^{1}$\quad INAF-IAPS, via del Fosso del Cavaliere 100, 00133 Rome, Italy; nicolas.deangelis@inaf.it\\
$^{2}$\quad Department of Enterprise Engineering ”Mario Lucenti”, University of Rome "Tor Vergata", Via Cracovia 50, 00133 Rome, Italy\\
$^{3}$\quad Department of Industrial, Electronic and Mechanical Engineering, "Roma Tre" University, via V. Volterra 62, 00146 Rome, Italy\\
$^{4}$\quad INAF-OAS Bologna, via Gobetti 93/3, 40129 Bologna, Italy\\
$^{5}$\quad INFN Sezione di Bologna, viale Berti Pichat 6/2, 40127 Bologna, Italy\\
$^{6}$\quad INAF-OAR, via Frascati 33, 00040 Monte Porzio Catone, Italy\\
$^{7}$\quad INAF-Headquarters, viale del Parco Mellini 84, 00136 Rome, Italy\\
$^{8}$\quad INAF-OACT, Via S. Sofia 78, 95123 Catania, Italy\\
$^{9}$\quad DEDA Connect s.r.l., via Vincenzo Lamaro 51, 00173 Rome, Italy\\
$^{10}$\quad DEIM, University of "La Tuscia", Largo dell’Università, 01100 Viterbo, Italy\\
$^{11}$\quad ASI, via del Politecnico snc, 00133 Rome, Italy\\
$^{12}$\quad IMT s.r.l., via Carlo Bartolomeo Piazza 30, 00161 Rome, Italy\\
$^{13}$\quad Department of Industrial Engineering, Alma Mater Studiorum Università di Bologna, Via Montaspro 97, 47121 Forlì, Italy\\
$^{14}$\quad Interdepartmental Centre for Industrial Aerospace Research, Alma Mater Studiorum Università di Bologna, Via Carnaccini 12, 47121 Forlì, Italy\\
$^{15}$\quad DIBAF, University of "La Tuscia", Largo dell’Università, 01100 Viterbo, Italy
}

\corres{Correspondence: nicolas.deangelis@inaf.it}

\abstract{Our Sun is the closest X-ray astrophysical source to Earth. As such, it makes a formidable case study to better understand astrophysical processes. Solar flares are in particular very interesting as they are linked to coronal mass ejections as well as magnetic field reconnection sites in the solar atmosphere. Flares can therefore provide insightful information on the physical processes at play on their production sites, but also on the emission and acceleration of energetic charged particles towards our planet, making it a formidable forecasting tool for space weather. While solar flares are critical to understanding magnetic reconnection and particle acceleration, their hard X-ray polarization -- key to distinguishing between competing theoretical models -- remains poorly constrained by existing observations. To address this, we present the CUbesat Solar Polarimeter (CUSP), a mission under development to perform solar flare polarimetry in the 25-100~keV energy range. CUSP consists of a 6U-XL platform hosting a dual-phase Compton polarimeter. The polarimeter is made of a central assembly of four 4$\times$4 arrays of plastic scintillators, each coupled to multi-anode photomultiplier tubes, surrounded by four strips of eight elongated GAGG scintillator bars coupled to avalanche photodiodes. Both types of sensors from Hamamatsu are respectively read out by the MAROC-3A and SKIROC-2A ASICs from Weeroc. In this manuscript, we present the preliminary spectral performances of single plastic and GAGG channels measured in the laboratory using development boards of the ASICs foreseen for the flight model.}

\keyword{Compton Polarimetry; X-ray Instrumentation; $\gamma$-ray Instrumentation; Calibration; Scintillators; APD; PMT; Solar Flares; Heliophysics; Space Weather; Magnetic Reconnection Processes} 

\begin{document}

\section{Solar Flare Polarimetry with CUSP}\label{sec:intro}

\subsection{The Physics of Solar Flares and the Need for Polarimetry}

Solar flares are among the most energetic phenomena in our solar system, arising from the sudden release of magnetic energy stored in the Sun’s corona. These explosive events, often associated with coronal mass ejections (CMEs), accelerate charged particles to relativistic speeds, heat plasma to tens of millions of Kelvin, and emit radiation across the electromagnetic spectrum -- from radio waves to gamma rays. The impulsive phase of solar flares, in particular, is dominated by hard X-ray (HXR) emission (typically above 20~keV), produced primarily by non-thermal bremsstrahlung from accelerated electrons interacting with the dense chromosphere \citep{Temmer2016, Nagasawa2022}.

Despite decades of multi-wavelength observations, key questions about the geometry of the magnetic reconnection region, the particle acceleration process, and the pitch-angle distribution and beaming of energetic electrons remain unanswered.

Spectroscopic and imaging observations alone often fail to break degeneracies between competing theoretical models. X-ray polarimetry provides a powerful, complementary diagnostic by probing the anisotropy of electron beams and the magnetic field geometry in the flare region \citep{Zharkova2010, Jeffrey2020}. The degree and orientation of linear polarization in HXRs are sensitive to the directionality of accelerated electrons (beamed vs. isotropic distributions), the local magnetic field configuration (e.g., loop geometry, reconnection site orientation), and the viewing angle relative to the flare site.

Theoretical models predict that non-thermal bremsstrahlung from beamed electrons can reach polarization fractions of tens of percent, while thermal bremsstrahlung (from hot, isotropic plasma) is only weakly polarized (a few percent at most) \citep{emslie1980, saint_hilaire2008}. Thus, polarimetric measurements in the 25–100~keV band can: \textit{(i)} distinguish thermal and non-thermal emission components in flares, \textit{(ii)} constrain the geometry of the acceleration region (e.g., loop-top vs. footpoint dominance), \textit{(iii)} test predictions of electron beaming and pitch-angle distributions.

To date, solar flare HXR polarimetry remains an under-explored frontier. Past measurements (e.g., from OSO-8, RHESSI, INTEGRAL, or CORONAS-F) have been statistically limited, often yielding only upper limits or marginal detections \citep{Tindo1970, Tindo1972a, tindo1972b, Tramiel1984, Boggs2006, SuarezGarcia2006, Zhitnik2006} even though some of these missions were dedicated polarimeters. Dedicated instruments with high sensitivity and time resolution are required to capture the rapid, impulsive evolution of flares. The CUbesat Solar Polarimeter (CUSP) mission is designed to address this gap by providing the first high-significance, time-resolved measurements of solar flare HXR polarization.

\subsection{Compton Polarimetry}

Polarimetry in the hard X-ray regime is challenging due to the penetrating nature of high-energy photons. Traditional polarimeters (e.g., Bragg crystals, photoelectric polarimeters) lose efficiency above $\sim$10 keV. Instead, Compton scattering polarimetry exploits the azimuthal asymmetry in the scattering cross-section of polarized photons.

The differential cross-section for Compton scattering is given by the Klein-Nishina formula \citep{KN_cross_section_paper}:

\begin{equation}
    \frac{d\sigma}{d\Omega} = \frac{r_e^2}{2} \left( \frac{E'}{E} \right)^2 \left[ \frac{E}{E'} + \frac{E'}{E} - 2 \sin^2 \theta \cos^2 \phi \right]
    \label{eq:klein_nishina}
\end{equation}

where $r_e$ is the classical electron radius (\(2.8 \times 10^{-15}\) m), \( E \) and \( E' \) are the incident and scattered photon energies, \( \theta \) is the polar scattering angle (angle between incident and scattered photon directions), and \( \phi \) is the azimuthal scattering angle (angle between the scattering plane and the polarization vector of the incident photon).

The $\cos^2 \phi$ dependence in Equation (\ref{eq:klein_nishina}) means that photons scatter preferentially perpendicular to their polarization vector. For an ensemble of photons, this results in an asymmetric azimuthal distribution of scattering angles (a.k.a. \textit{modulation curve}) in the case of a polarized flux. The relative amplitude of this modulation is directly related to the polarization degree (PD) of the source, while the phase of the modulation gives the polarization angle (PA).

The modulation factor ($\mu_{100}$) quantifies the instrument’s sensitivity to polarization:

\begin{equation}
    \mu_{100}(\theta) = \frac{N_{\text{max}}(\theta) - N_{\text{min}}(\theta)}{N_{\text{max}}(\theta) + N_{\text{min}}(\theta)} = \frac{\sin^2 \theta}{\frac{E}{E'} + \frac{E'}{E} - \sin^2 \theta}
    \label{eq:modulation_factor}
\end{equation}

where \( N_{\text{max}} \) and \( N_{\text{min}} \) are the maximum and minimum counts in the azimuthal distribution. For $E \ll 511$ keV (the electron rest-mass energy), the maximum modulation occurs at $\theta = 90^\circ$ (orthogonal scattering), where \(\mu_{100} \approx 1\) (100\% modulation for 100\% polarized light).

A figure of merit of the sensitivity of polarimeters, the Minimum Detectable Polarization (MDP) at 99\% confidence, is given by \citep{Weisskopf2010}:

\begin{equation}
    \text{MDP}_{99\%} = \frac{4.29}{\mu_{100} \sqrt{R}} \sqrt{\frac{R + B}{T}}
    \label{eq:mdp}
\end{equation}

where $R$ and $B$ are the source and background count rates, and $T$ is the observation time.

A high modulation factor and efficient detection are thus critical for achieving low MDP values and being sensitive to lowly polarized signals.

\subsection{The CUSP Polarimeter: Instrument Design and Principle of Operation}

\begin{figure}[htpb]
\centering
 \includegraphics[height=.35\textwidth]{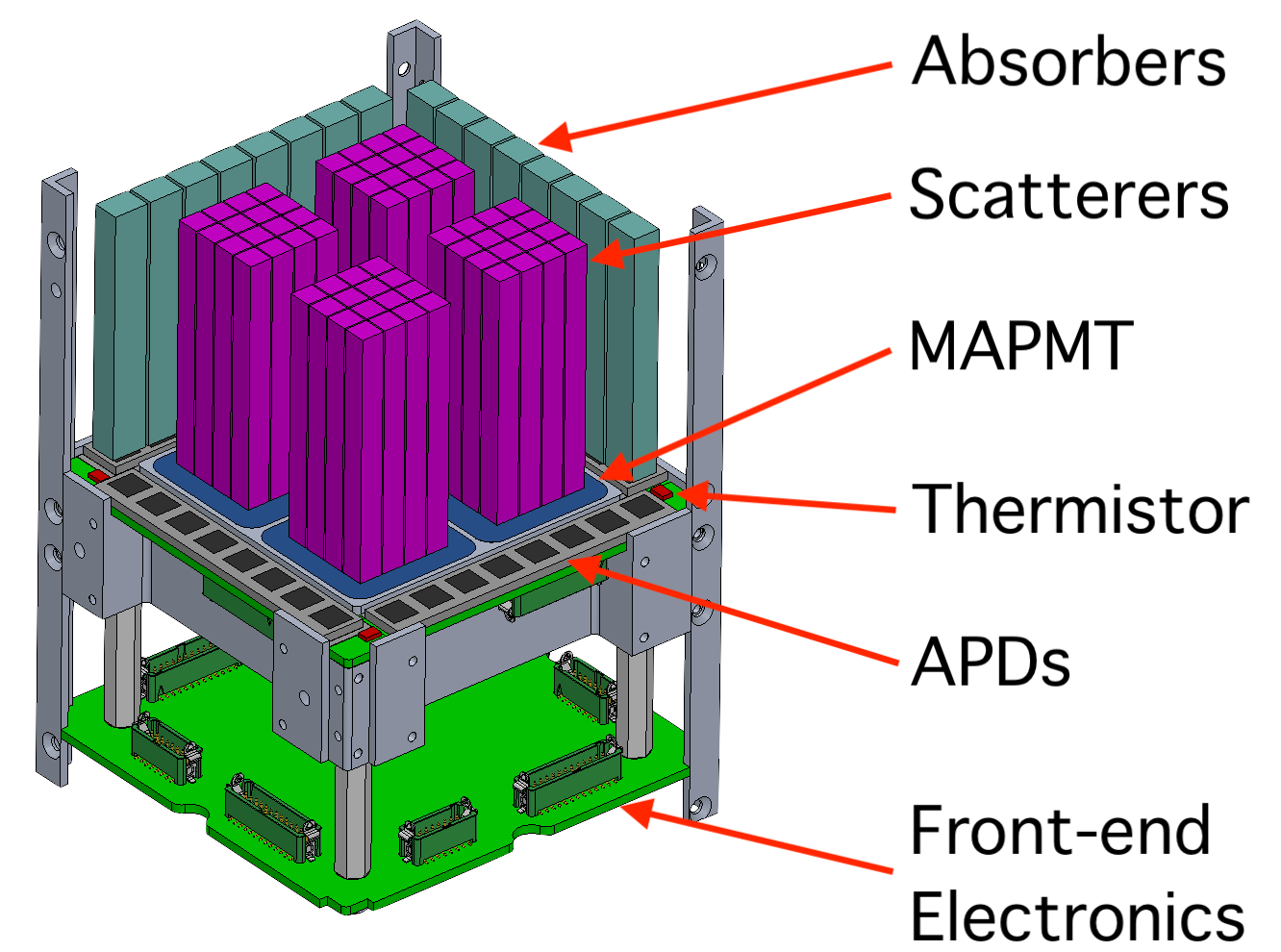}\hspace*{0.5cm}\includegraphics[height=.42\textwidth]{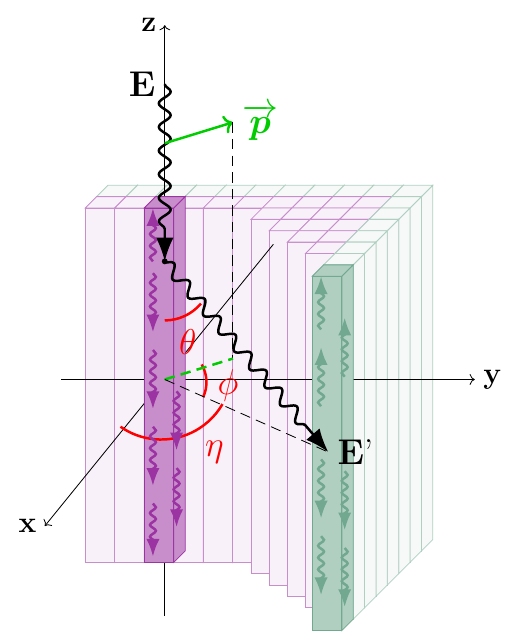}
 \caption{\textbf{Left:} Partial CAD design of CUSP's hard X-ray polarimeter showing its sensitive parts. \textbf{Right:} Schematic working principle of a Compton polarimeter. The incoming photon is Compton scattered in a segment of the detector, depositing some energy which is converted into scintillation optical light and collected by a photosensor at the extremity of the scintillator bar. The scattered photon is then absorbed in a different segment of the instrument, which allows for determining the azimuthal scattering direction of the primary photon.}
 \label{fig:cusp_design}
\end{figure}

The CUbesat Solar Polarimeter (CUSP) is a 6U-XL CubeSat mission developed under the Italian Space Agency’s (ASI) Alcor program, designed to measure the linear polarization of solar flares in the 25–100~keV band using a dual-phase Compton polarimeter. The instrument, whose CAD design is shown in Figure \ref{fig:cusp_design} together with its detection working principle, consists of a scatterer array of 64 plastic scintillator bars (low-Z material to maximize Compton scattering) read out by four 16-channel Multi-Anode Photomultiplier Tubes (MAPMTs), surrounded by a ring of absorbers made of 32 Cerium-doped GAGG (Gd$_3$Al$_2$Ga$_3$O$_{12}$:Ce) scintillator crystals (high-Z material to maximize photoelectric absorption) read out by Avalanche Photodiodes (APDs). The R7600-03-M16-Y002 Ultra Bialkali MAPMT and the S16554-55S APD, both from Hamamatsu, are respectively read out by the MAROC-3A and SKIROC-2A ASICs from Weeroc. The baseline for the scatterer is Eljen's PVT-based EJ-204 scintillator, although other materials from the same family are being investigated.

When an incoming HXR photon interacts in the scatterer, it Compton-scatters, depositing partial energy before being absorbed in a GAGG scintillator, as depicted in Figure \ref{fig:cusp_design}. The azimuthal scattering angle ($\phi$) is reconstructed from the relative positions of the scatter and absorption events. By accumulating many such events, the modulation curve is built, from which the polarization degree (PD) and angle (PA) are derived. A tungsten collimator reduces the field of view of the polarimeter to $\pm 36^\circ$, and Al/Ti and Al/Ti/W filters are respectively used for the scatterers and absorbers to filter soft X-rays and electrons.

Simulations of benchmark solar flares from \citep{saint_hilaire2008} have shown that CUSP will have a 3.9\% MDP for an X1.2 flare with a duration of 240~s. The MDP requirement for CUSP is of 5\% in the 25-100~keV range for an X1.2 solar flare with a observation time of 240~s. More details on CUSP's scientific prospects and mission concept can be found in \citep{SPIE25_Fabiani_CUSP}.

%\textcolor{red}{Table \ref{tab:mdp_req} gives the requirements on the MDP for four different classes of flares.}

%\begin{table}[ht]
%\centering
%\begin{tabular}{lccr}
%\toprule
%Flare class & MDP$_{99\%}$ in 25-100~keV & Flare duration (s) & Benchmark flare ID \\ \midrule
%M1 & 72.0 & 128 & 20020927-03\_34\_11-03\_36\_19 \\
%M5.2 & 7.8 & 284 & 20050822-17\_01\_53-17\_06\_38 \\
%X1.2 & 3.9 & 240 & 20030529-01\_00\_59-01\_04\_59 \\
%X10 & 0.05 & 351 & 20031029-20\_40\_15-20\_46\_06 \\ \bottomrule
%\end{tabular}
%\caption{\textcolor{red}{Minimum Detectable Polarization requirements defined for various classes of flares defined based on benchmark flares.}}
%\label{tab:mdp_req}
%\end{table}

In order to validate the readout chain and spectral performances of both the scatterers and absorbers of CUSP's polarimeter, single-channel setups based on both types of detectors read out by ASIC development boards were assembled. While preliminary results obtained with this setup were reported in \citep{SPIE25_DeAngelis_CUSP}, we describe here the detailed spectral performances as measured in the laboratory for both detection chains.

\section{Single Channel Spectral Performances}\label{sec:results}
\subsection{Scatterers}\label{subsec:scatterers}

\begin{figure}[htpb]
\centering
 \includegraphics[height=.42\textwidth]{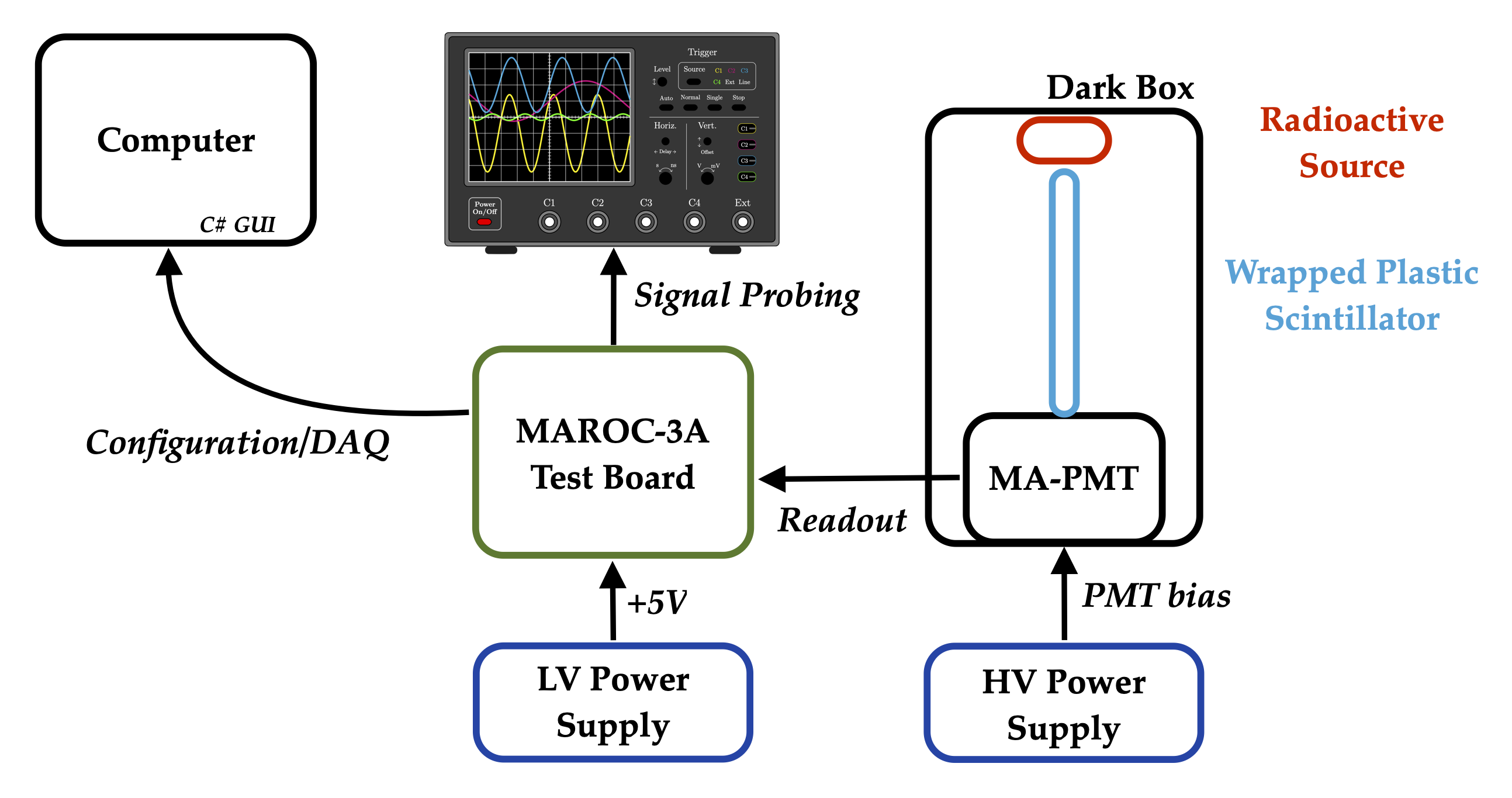}\includegraphics[height=.42\textwidth]{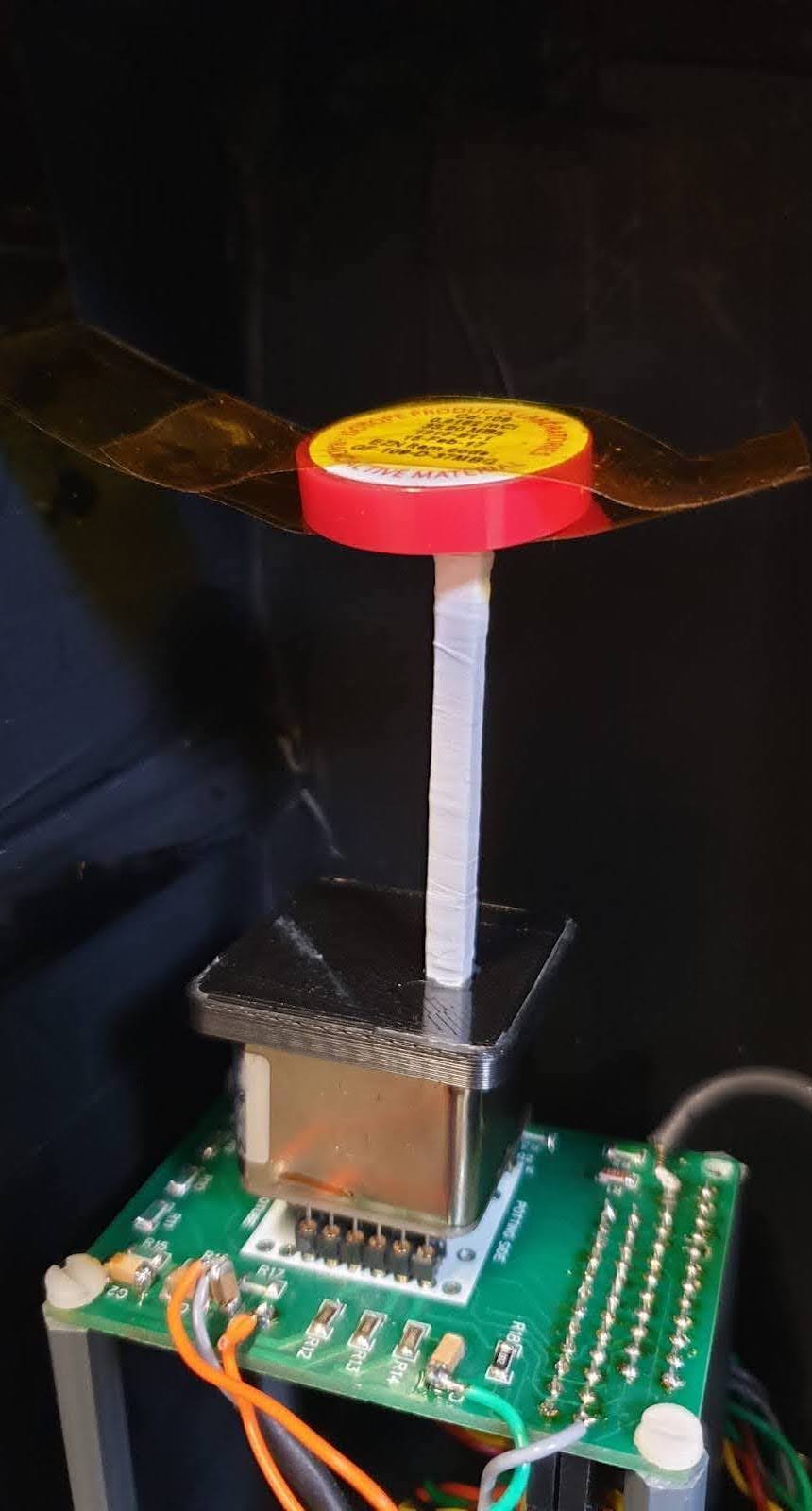}
 \caption{\textbf{Left:} Diagram showing the laboratory setup used to characterize the CUSP scatterer. The MAROC-3A development board is reading out the MA-PMT used to read out the plastic scatterer. \textbf{Right:} Plastic scintillator bar wrapped using PTFE tape and coupled to an MA-PMT channel. A $^{109}$Cd source is placed on top of the scintillator and the assembly is placed in a dark box.}
 \label{fig:scatterers_setup}
\end{figure}

The setup for testing a single-channel scatterer is based on a plastic scintillator bar wrapped with several layers of 100~$\mu$m-thick polytetrafluoroethylene (PTFE; a.k.a. Teflon) tape and coupled to a channel of an R7600-03-M16-Y002 UBA MAPMT from Hamamatsu using a custom mechanical jig and optical grease. The assembly is placed in a light-tight box and the MAPMT channels are read by a MAROC-3A development board connected to a computer equipped with a C\# GUI software used for ASIC configuration and data acquisition. A schematic representation of the setup as well as a picture of the detector assembly inside the dark box are shown in Figure \ref{fig:scatterers_setup}.

\begin{figure}[htpb]
\centering
 \hspace*{-0.5cm}\includegraphics[height=.4\textwidth]{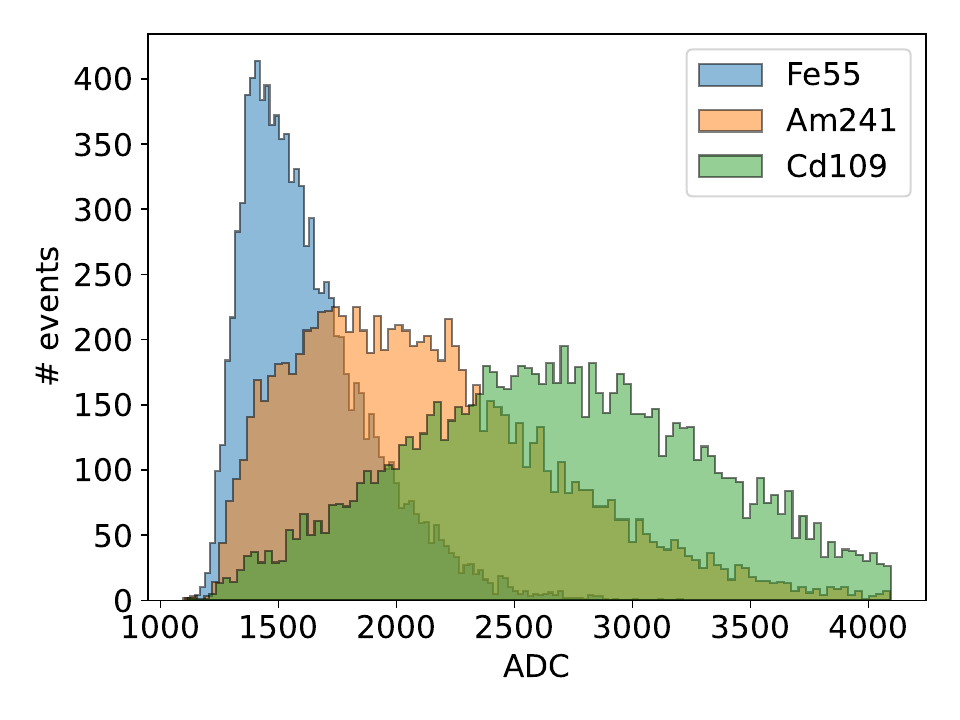}\includegraphics[height=.4\textwidth]{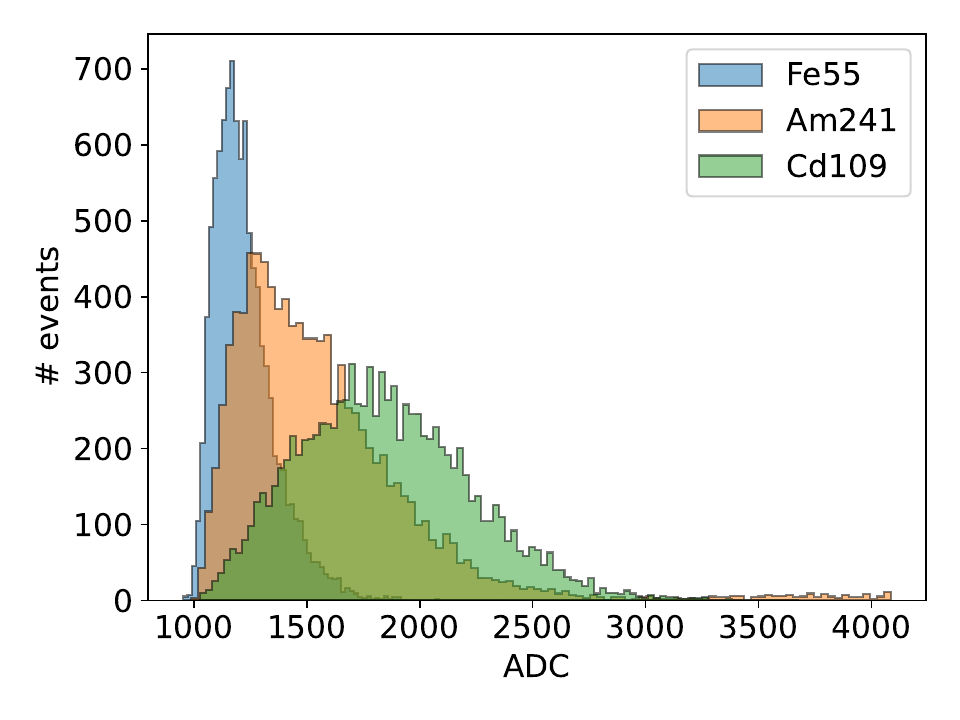}
 \caption{\textbf{Left:} $^{55}$Fe, $^{109}$Cd, and $^{241}$Am spectra measured with an EJ-204 scintillator bar using the unipolar fast shaper, a preamplifier gain of 0.2, and a bias voltage of 840~V. \textbf{Right:} Same spectra measured using MAROC-3A's bipolar fast shaper 1 at a bias of 820~V and with a preamplifier gain of 0.15.}
 \label{fig:plastic_spectra_sources}
\end{figure}

Spectra were acquired by placing various radioactive isotopes with characteristic features in the energy range of interest on top of the plastic scintillator. Three different signal shapers are available for triggering with the MAROC-3A; namely, a unipolar fast shaper (fsu), a bipolar fast shaper (fsb1), and a half bipolar fast shaper (fsb2) \citep{maroc3a_datasheet}. The unipolar fast shaper allows triggering on weak signals, with a charge down to 50~fC or lower, while the half bipolar fast shaper is suitable for very strong signals. Both the 'fsu' and 'fsb1' are being investigated here, although the unipolar fast shaper is expected to be the most suitable for the scatterer signals as a small amount of optical light is expected from the scintillation in plastic scintillators. A slow shaper, with shaper peaking time in the range of 50-142~ns, is used to read the charge signal. The measured $^{55}$Fe, $^{241}$Am, and $^{109}$Cd spectra using both 'fsu' and 'fsb1' fast shapers for triggering are plotted in Figure \ref{fig:plastic_spectra_sources}. The 5.95~keV line of $^{55}$Fe, a blend of lines around 18.0~keV from $^{241}$Am (as detailed in section \ref{subsec:absorbers}), and a blend of lines around 23.108~keV from $^{109}$Cd, can be observed in both cases. $^{109}$Cd has a K$_{\alpha2}$ line at 22.984~keV with 4.31\% branching ratio and a K$_{\alpha1}$ line at 23.174~keV with a branching ratio of 8.12\% \citep{NuDat}, which gives a weighted energy of 23.108~keV.

\begin{figure}[htpb]
\centering
 \hspace*{-0.5cm}\includegraphics[height=.4\textwidth]{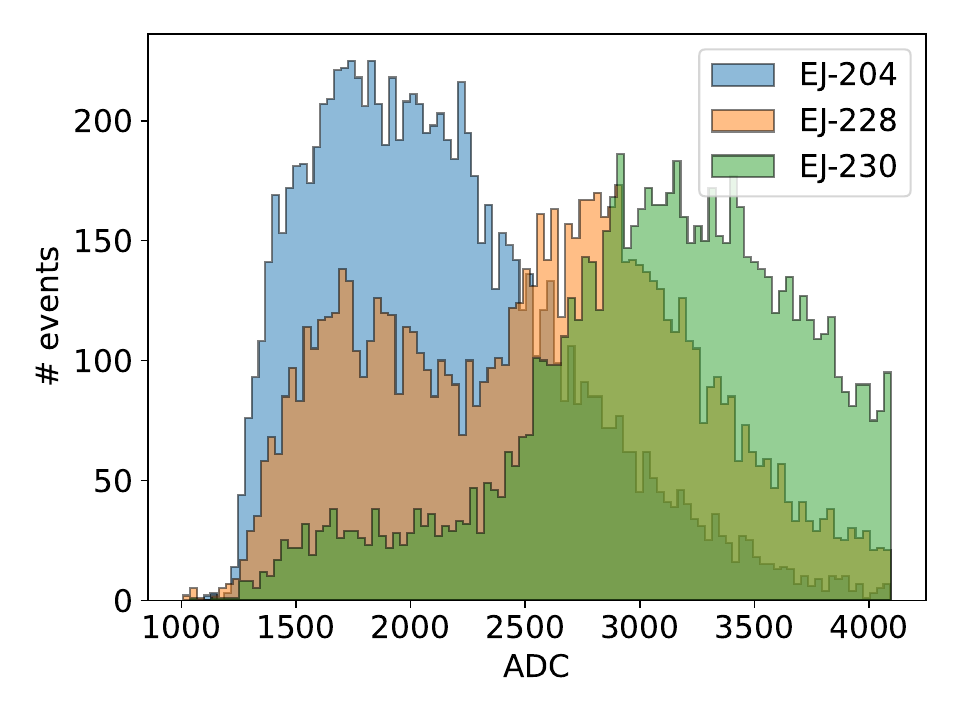}\includegraphics[height=.4\textwidth]{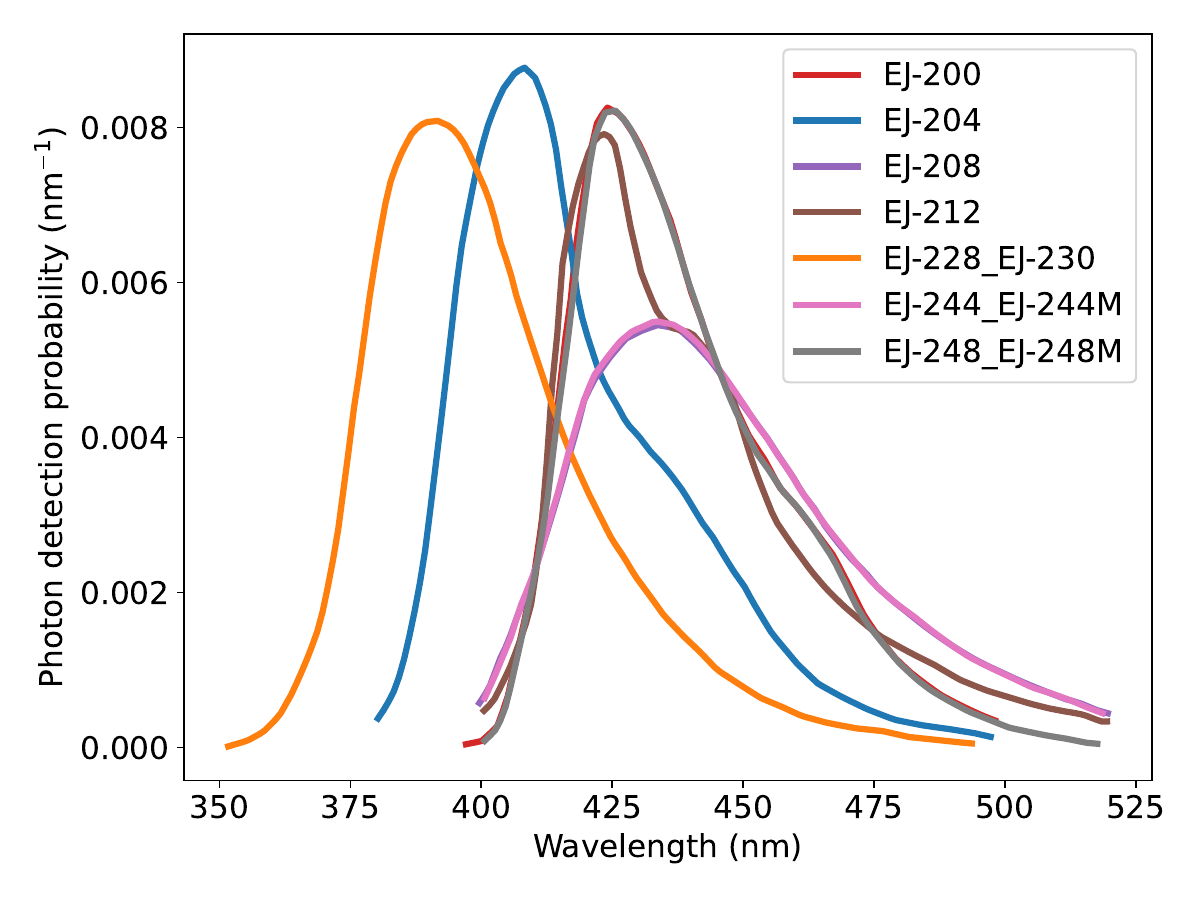}
 \caption{\textbf{Left:} $^{241}$Am spectra measured using the unipolar fast shaper with a bias of 840~V with three kinds of plastic scintillators. \textbf{Right:} Scintillation spectra of various PVT-based scintillators convolved with the quantum efficiency of the R7600-03-M16-Y002 MAPMT from Hamamatsu.}
 \label{fig:Am241_spectra_splasticscintchoice}
\end{figure}

Eljen's EJ-204 plastic scintillator\footnote{Eljen EJ-204 -- https://eljentechnology.com/products/plastic-scintillators/ej-200-ej-204-ej-208-ej-212 -- \textit{Consulted on 19 September 2025.}} is used as the baseline for the scattering material because of its high scintillation efficiency of 10400 photons/MeV. Other scintillators such as EJ-228 or EJ-230 have a slightly lower scintillation efficiency but a better spectral compatibility with the quantum efficiency of the MAPMT. They are therefore also being investigated as potential scattering materials. The spectral compatibility of scintillators with the sensor, i.e. between the emitted spectra from the scintillators and the quantum efficiency spectrum of the sensor, is computed by multiplying the scintillator and MAPMT spectra, resulting into a photon detection probability density spectrum shown in Figure \ref{fig:Am241_spectra_splasticscintchoice}. This figure also shows the spectra obtained for the three types of scintillators mentioned earlier with an $^{241}$Am source, where EJ-228 and EJ-230 appear to allow a more efficient light collection despite their slightly lower scintillation efficiency.

Another crucial consideration for the selection of the plastic scintillator material, except for the scintillation efficiency and spectral compatibility to the MAPMT, is the light output fraction that one gets out of the scintillator bar. That is the fraction of scintillation light that actually gets to the entrance of the detector after interface losses. This fraction can be affected by the shape of the scintillators, but more importantly by the roughness of the scintillator's polished surfaces which is directly linked to the softness of the material \citep{De_Angelis_2025_optsim}. Table \ref{tab:plastic_scint_light_output} gives the integrated spectral compatibility for various plastic scintillators from Eljen in detected photons (or photoelectrons) per scintillation photon, as well as the scintillation efficiency of the material and the light yield of the scintillator in detected photons per unit of deposited energy assuming a light output fraction of 0.3 for all scintillators. This fraction varies depending on the scintillator surface roughness which can change the photon losses between the scintillator and the sensor, causing a material with higher scintillation efficiency to lead to a lower light yield due to rougher surfaces \citep{De_Angelis_2025_optsim}. Many of the plastic scintillator materials quoted in Table \ref{tab:plastic_scint_light_output} will therefore be investigated in the future to determine which material offers an optimal light collection for our design.

\begin{table}[htpb]
\centering
\begin{tabular}{lccc}
\toprule
Scintillator & Spectral Compatibility & Scintillation Efficiency & Light Yield \\
Material & (p.e./opt.ph.) & (opt.ph./1 MeV e$^-$) & (p.e./keV) \\ \midrule
EJ-200 & 0.35 & 10'000 & 1.036 \\
EJ-204 & 0.37 & 10'400 & 1.142 \\
EJ-208 & 0.33 & 9'200 & 0.913 \\
EJ-212 & 0.34 & 10'000 & 1.019 \\
EJ-228 & 0.38 & 10'200 & 1.152 \\
EJ-230 & 0.38 & 9'700 & 1.096 \\
EJ-244(M) & 0.33 & 8'600 & 0.854 \\
EJ-248(M) & 0.34 & 9'200 & 0.950 \\ \bottomrule
\end{tabular}
\caption{Comparison of the integrated spectral response of various PVT-based scintillators from Eljen. The light yield, i.e. the amount of optical light detected per unit of deposited energy in the scintillator, is computed assuming a scintillator light output fraction of 0.3, that is, only 30\% of the optical light is collected by the sensor. This number varies depending on the surface roughness obtained via diamond polishing, which in turn depends on the material itself \citep{De_Angelis_2025_optsim}.}
\label{tab:plastic_scint_light_output}
\end{table}

\begin{figure}[htpb]
\centering
 \hspace*{-0.5cm}\includegraphics[height=.4\textwidth]{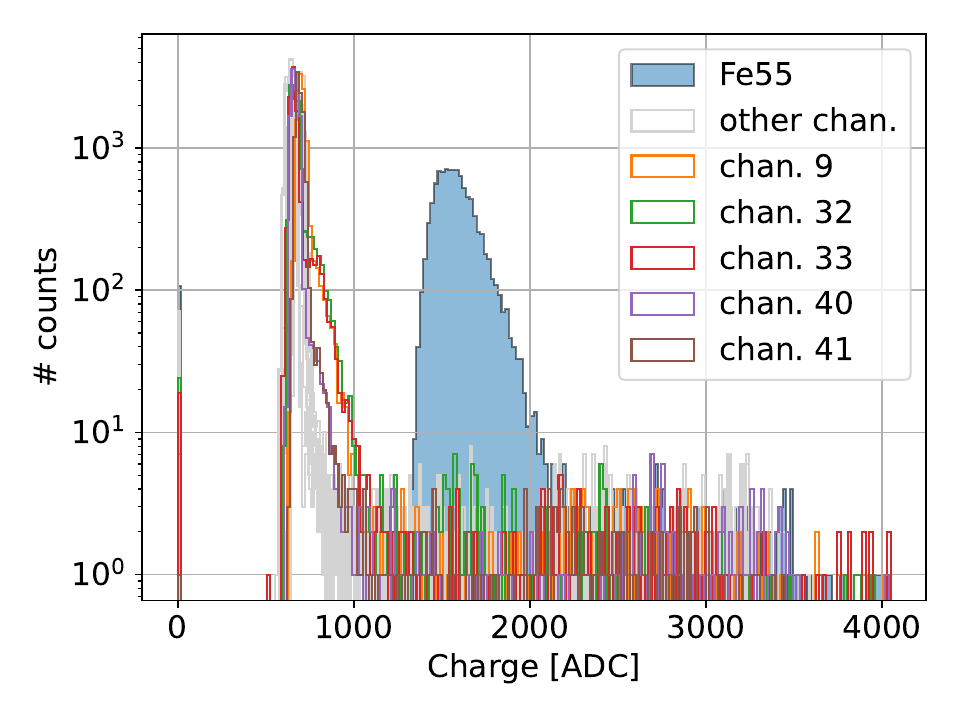}\includegraphics[height=.4\textwidth]{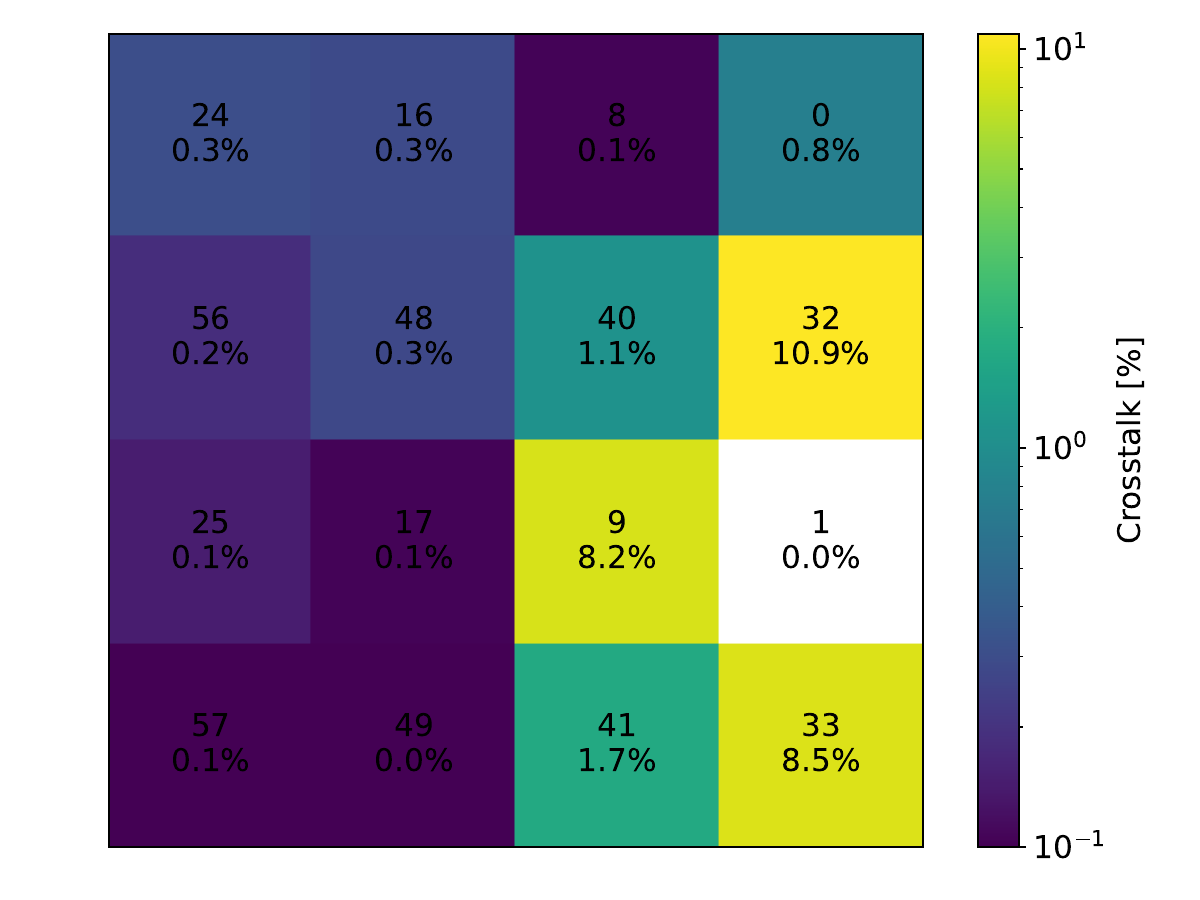}
 \caption{\textbf{Left:} Spectra measured by the 16 MAPMT channels, with channel 1 coupled to a wrapped plastic scintillator with an $^{55}$Fe on top, and all other channels covered by an opaque jig (see Figure~\ref{fig:scatterers_setup}, right). The effect of optical crosstalk through the MAPMT's entrance window can be seen in the spectra of the channels neighboring channel 1. \textbf{Right:} Crosstalk map showing an optical crosstalk of up to 10.9~\% in the direct neighbors and non-significant crosstalk to non-neighboring channels. The channel ID and crosstalk value are indicated on each bin.}
 \label{fig:plastic_crosstalk}
\end{figure}

Another important quantity impacting the instrument's response is the optical crosstalk, or the amount of light that gets detected in channels neighboring the one where the actual event is occurring. An example of a spectrum collected with the single wrapped plastic bar coupled to channel 1 is shown together with the spectra of the other 15 channels in Figure \ref{fig:plastic_crosstalk}. An excess in the spectrum above the dark noise peak can be observed for some channels that are neighbors of channel 1. From these spectra, one can extract the fraction of light leaking from channel 1 to its neighbors and compute an optical crosstalk map, also shown in Figure \ref{fig:plastic_crosstalk}. Most of the channels show a negligible amount of crosstalk (within statistical fluctuations), except for their direct and diagonal neighbors, for which the optical crosstalk is respectively comprised in the ranges of 8-11\% and 1-2\%.

A more accurate value of the crosstalk can be obtained using a complete array of plastic scintillators, more representative than a single plastic scintillator coupled to an MAPMT channel. However, the main contribution to the optical crosstalk is due to the 0.8~mm-thick entrance 'UV-glass' of the MAPMT, and the crosstalk estimates given here are therefore a good approximation. The values obtained also match the optical crosstalk of POLAR, a GRB polarimeter using similar MAPMTs, for which the crosstalk was in the range of 10 to 15\% \citep{De_Angelis_2025_optsim}. The crosstalk asymmetry in the vertical direction of the map is likely due to a small misalignment of the bar compared to the center of the MAPMT channel, which will be easily fixed in the future when using dedicated mechanics encapsulating the wrapped scintillator array. The optical crosstalk impacts the instrument's energy threshold as part of the optical light from the channel in which the X-ray is interacting will leak to the neighbor channels, and should therefore be minimized. The optical crosstalk should also be fully described by simulations combined with laboratory measurements in order to include the induced systematic effects in both spectral and polarization response matrices.

\begin{figure}[htpb]
\centering
 \hspace*{-0.5cm}\includegraphics[height=.4\textwidth]{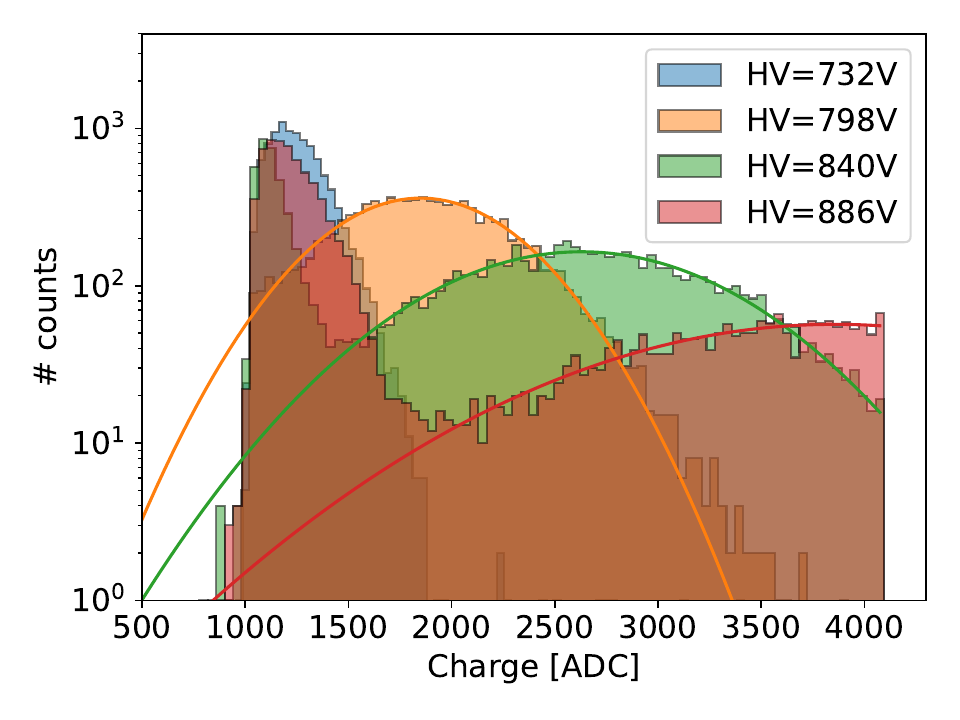}\includegraphics[height=.4\textwidth]{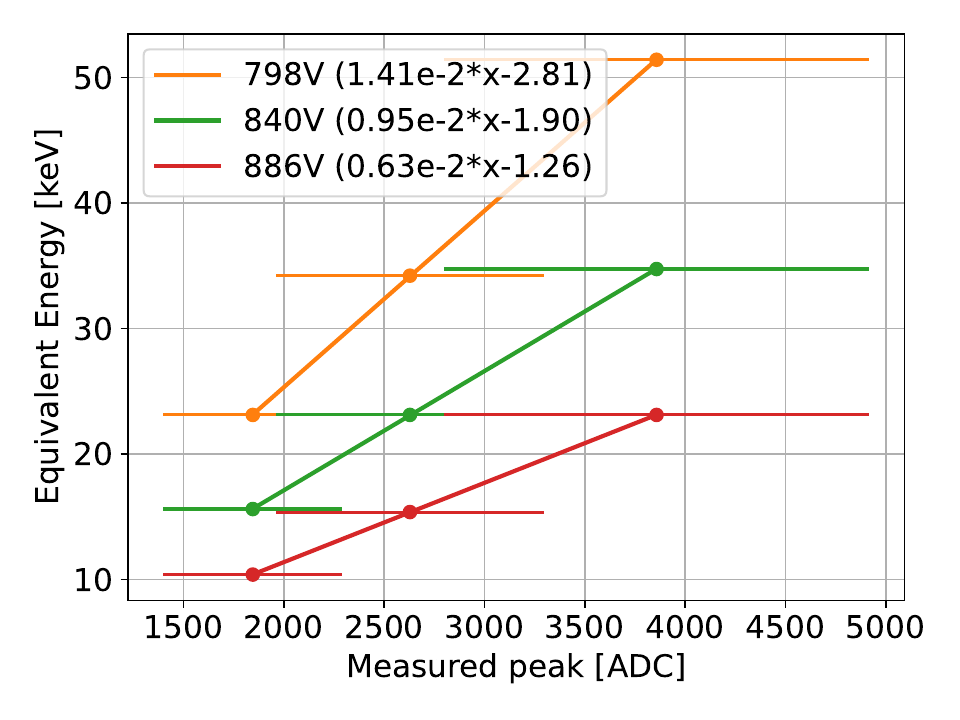}
 \caption{\textbf{Left:} $^{109}$Cd spectra acquired for various PMT bias voltages. \textbf{Right:} Equivalent energy of the 23.108~keV $^{109}$Cd feature (two lines weighted, see explanations in the text) assuming a given voltage, and applying a gain ratio based on the PMT's gain dependence on the high voltage, as a function of the photopeak position.}
 \label{fig:cd109_dynamicrange_fit}
\end{figure}

For a Compton scattering event for which the polar scattering angle is 90$^\circ$, i.e. for which the Compton energy deposition is maximal, the energies deposited through Compton scattering by a 25 and 100~keV incoming photon are respectively 1.17 and 16.4~keV. The 25-100~keV energy range of CUSP therefore translates into a Compton deposition in the plastic scatterers in the range of 1.17--16.4~keV. The dynamic range of the scatterer can be estimated using the $^{109}$Cd spectrum, fitted in Figure~\ref{fig:cd109_dynamicrange_fit}. One can rescale the 23.108~keV photopeak position (from the two lines mentioned earlier) in ADC for each bias voltage by the PMT gain ratio \citep{PMTdatasheet} between the different bias voltages applied to get an idea of the energy (keV) vs charge (ADC) for all three high voltages, as plotted in Figure~\ref{fig:cd109_dynamicrange_fit}. The 23.108~keV peak position can be used to determine the energy dynamic range that can be achieved with the absorber's acquisition chain based on development boards.

\begin{figure}[htpb]
\centering
 \hspace*{-0.5cm}\includegraphics[height=.4\textwidth]{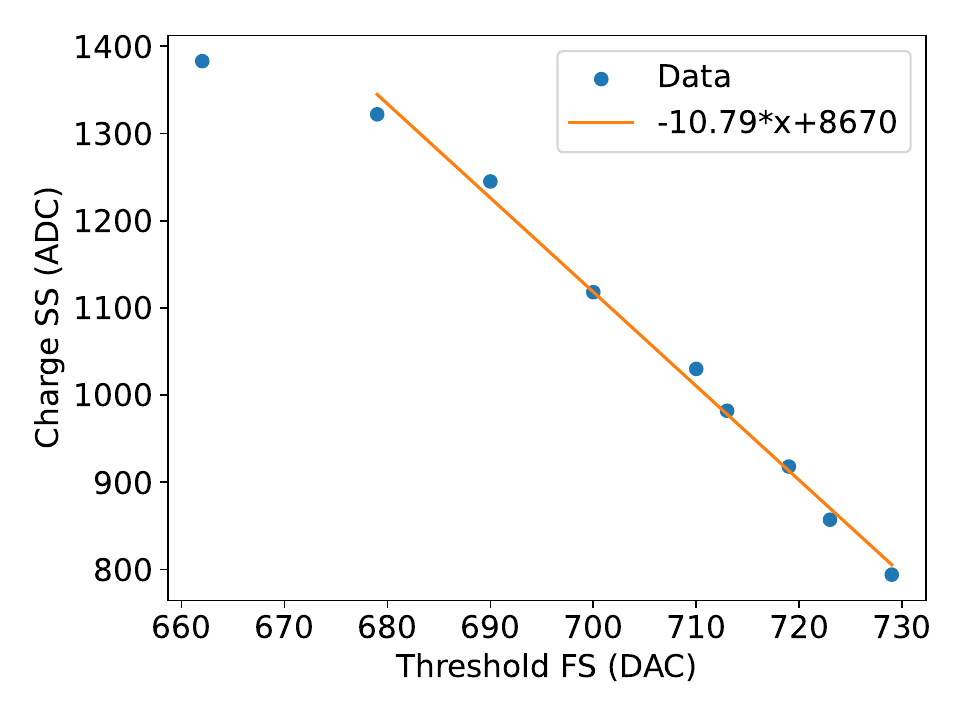}\includegraphics[height=.4\textwidth]{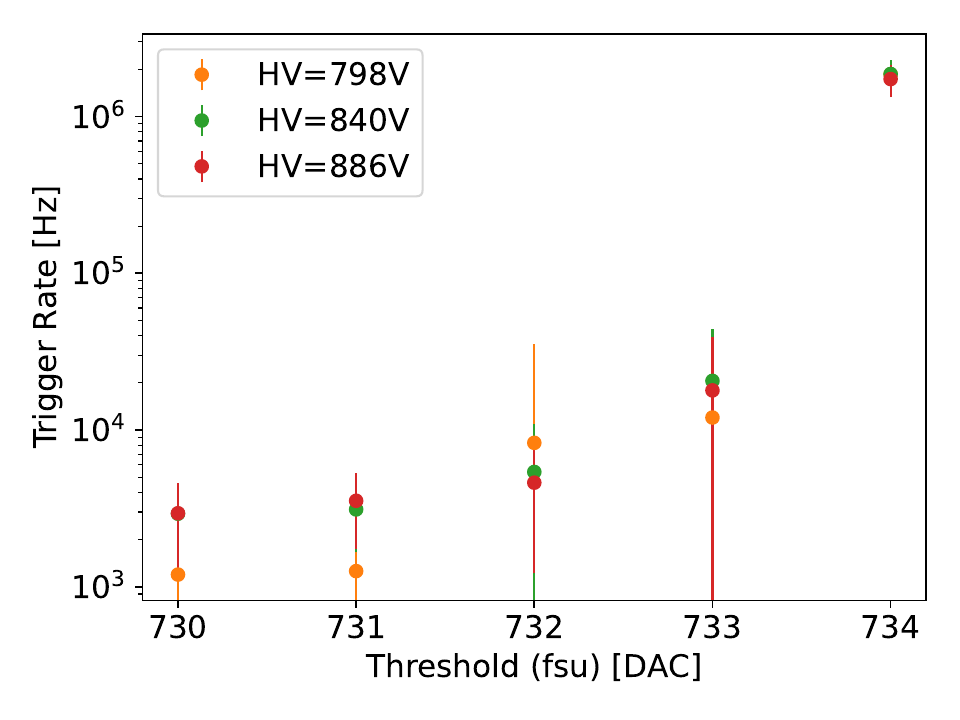}
 \caption{\textbf{Left:} Conversion of the digital threshold value into digitized charge. It should be noted that the threshold scale is inverted, i.e. a lower DAC value corresponds to a higher threshold. \textbf{Right:} Noise trigger rate as a function of the threshold value for the unipolar fast shaper for different PMT bias voltages.}
 \label{fig:threshold_rate}
\end{figure}

The conversion between measured charge and threshold digital value is shown in Figure~\ref{fig:threshold_rate} together with the noise rate as a function of the threshold value. One should note that the threshold scale is inverted, meaning that a higher DAC value implies a lower voltage. From the noise rate measured at various bias voltages, the minimum threshold value that can be used with the development board setup is 733~DAC, for which the noise rate is still easily manageable ($\ll$1~MHz) once applying a Compton coincidence window of the order of hundreds of nanoseconds. This threshold value is far from optimized, as the ASIC development board was designed by Weeroc for functional tests, thus not offering the best achievable performance. This setup is therefore quite noisy, and a much lower threshold value should easily be achieved using custom electronics in the future. The 733~DAC threshold corresponds to a charge of 762~ADC, which is considered hereafter as the energy threshold of the scatterer. The upper energy boundary is computed as the energy at 4096~ADC. Using this charge range and the $^{109}$Cd photopeak position to convert the charge into energy, one can compute the energy range of the absorber as a function of the PMT bias voltage, as given in Table~\ref{tab:energy_range}.

At a voltage of 886~V, the achieved dynamic range is 3.55--24.62~keV, neglecting ASIC saturation effects which could affect the linearity at high energy as well as the upper energy bound. Although the targeted energy range for Compton depositions is 1.17--16.4~keV and is therefore not achieved with this setup, the use of custom electronics optimized for noise reduction will significantly improve the performance at low energies and allow for reducing the energy threshold. A prototype with dedicated electronics rather than development boards is currently under development as part of CUSP's phase B, as described in Section~\ref{subsec:coincidence}.

\begin{table}[htpb]
\centering
\begin{tabular}{lccc}
\toprule
PMT bias voltage (V) & 798 & 840 & 886 \\ \midrule
%PMT gain \citep{PMTdatasheet} & 1991415 & 2947272 & 4429856 \\
PMT gain ($10^6$) \citep{PMTdatasheet} & 1.99$\pm$0.05 & 2.95$\pm$0.05 & 4.43$\pm$0.05 \\
\textcolor{red}{Energy (keV) calibrated with ref. HV=798V} & \textcolor{gray!80}{23.108} & 34.20 & 51.40 \\
\textcolor{red}{Energy (keV) calibrated with ref. HV=840V} & 15.61 & \textcolor{gray!80}{23.108} & 34.73 \\
\textcolor{red}{Energy (keV) calibrated with ref. HV=886V} & 10.39 & 15.37 & \textcolor{gray!80}{23.108} \\
Measured peak (ADC) & $1846 \pm 446$ & $2629 \pm 667$ & $3858 \pm 1060$ \\
Energy threshold (762~ADC) (keV) & 7.91 & 5.34 & 3.55 \\
Upper energy boundary (4'096~ADC) (keV) & 54.77 & 37.01 & 24.62 \\
\bottomrule
\end{tabular}
\caption{Determination of the energy dynamic range for three values of PMT bias voltages. The PMT gain values have been obtained through the fit of the gain versus high voltage plot from the Hamamatsu datasheet \citep{PMTdatasheet}.}
\label{tab:energy_range}
\end{table}

\subsection{Absorbers}\label{subsec:absorbers}

\begin{figure}[htpb]
\centering
 \includegraphics[height=.42\textwidth]{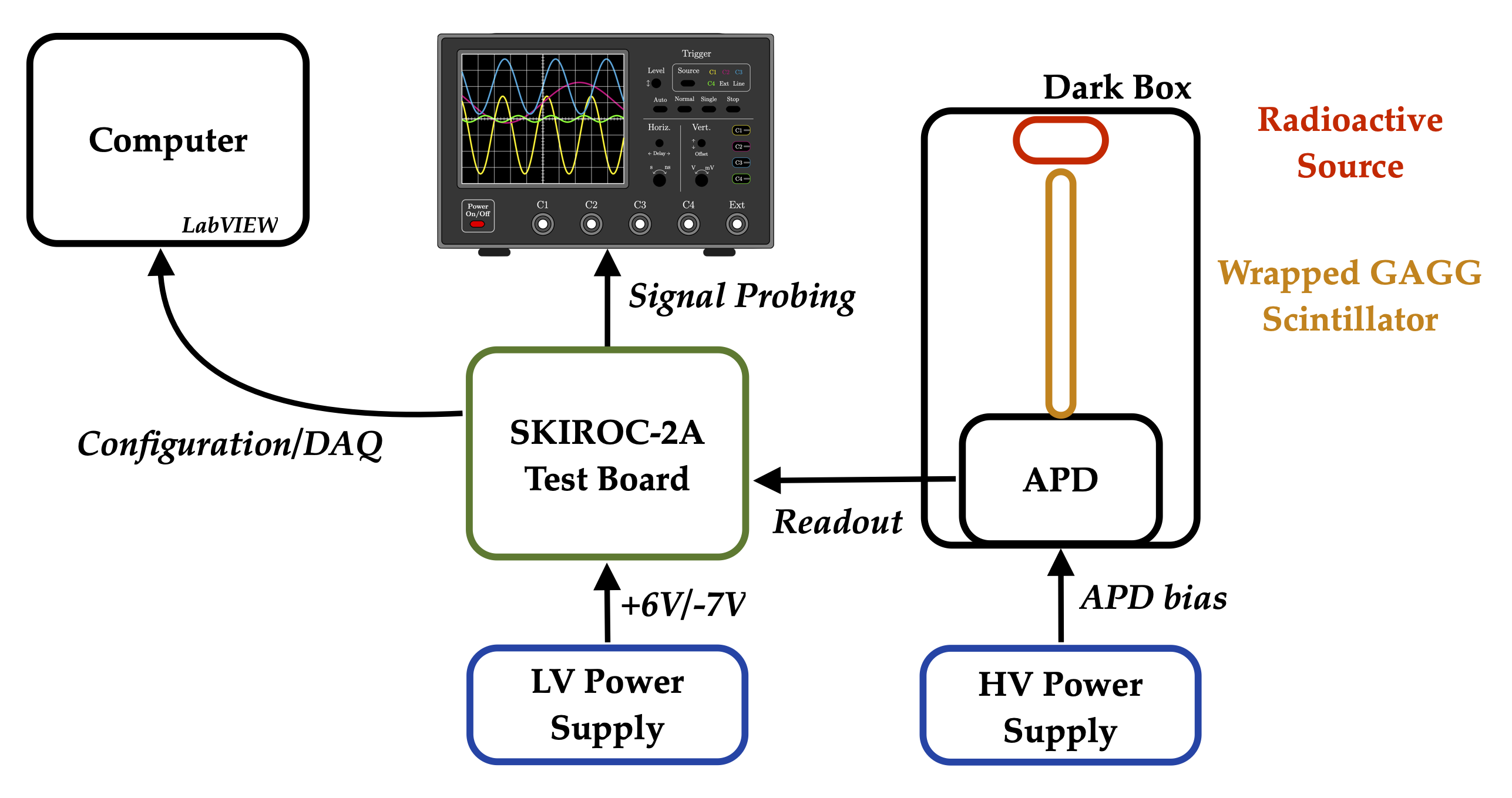}\includegraphics[height=.42\textwidth]{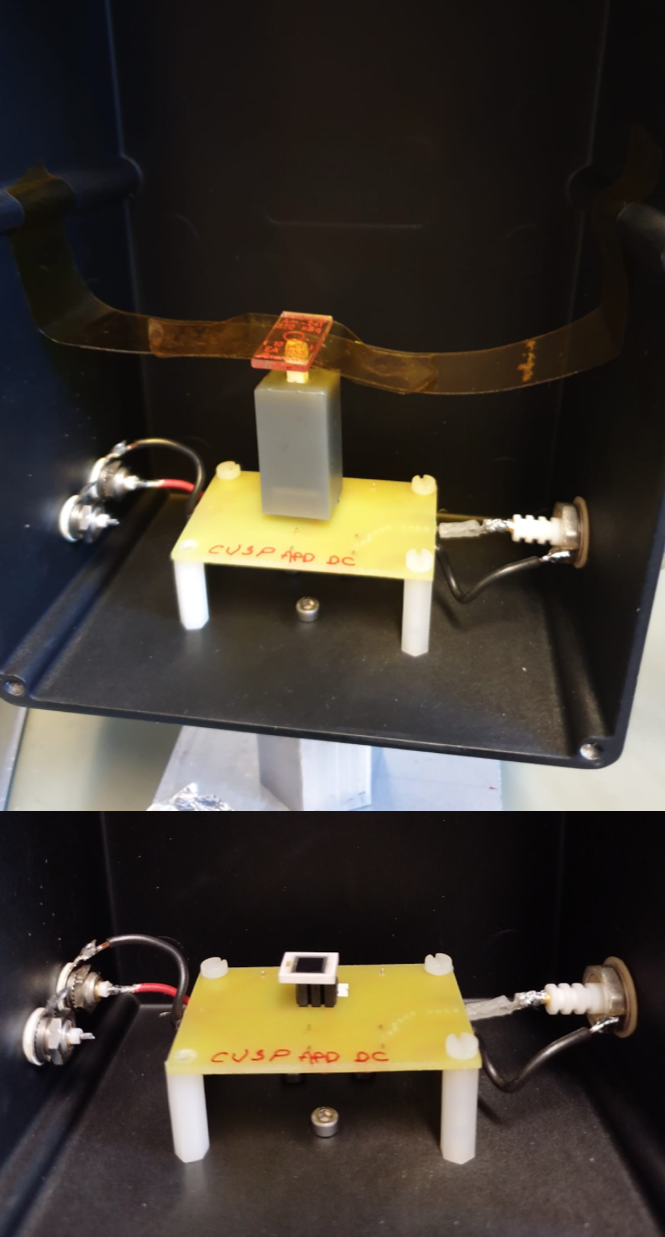}
 \caption{\textbf{Left:} Diagram showing the laboratory setup used to characterize the CUSP absorber. The SKIROC-2A development board is reading out the APD coupled to a GAGG absorber. \textbf{Right:} APD mounted on its RC filter board (bottom) and coupled to a wrapped GAGG scintillator bar on which an $^{241}$Am source is placed (top).}
 \label{fig:absorbers_setup}
\end{figure}

A setup similar to that use for the scatterers has been developed to test the absorber acquisition chain. It is based on a GAGG scintillator bar wrapped in PTFE tape coupled to a single-channel S8664-55 APD from Hamamatsu \citep{APDdatasheet}. This sensor has a ceramic package with pins and an epoxy resin, while the flight APD S16554-55S has an SMD package and a silicone resin. The sensor itself is however the same. The detector assembly is once again placed in a light-tight box, and a SKIROC-2A development board, this time controlled via LabVIEW software, is used to acquire the signal. Both the diagram of the setup and a picture of the single-channel detector assembly are shown in Figure \ref{fig:absorbers_setup}. Only the analog part of the ASIC is used for limiting the power consumption of the system, and because the digital part does not allow a fast enough event readout for our application due to high dead-time \citep{skiroc2a_datasheet}.

\begin{figure}[htpb]
\centering
 \hspace*{-0.5cm}\includegraphics[height=.4\textwidth]{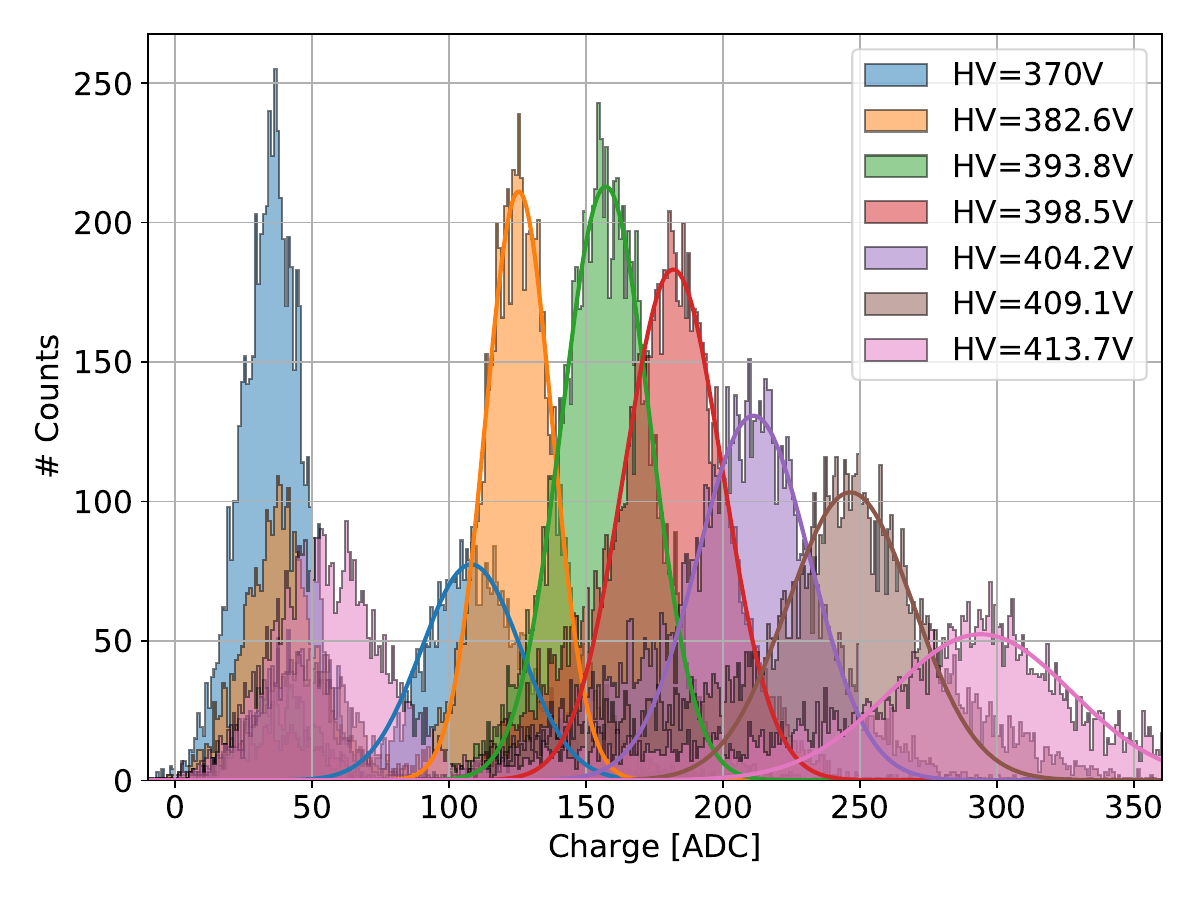}\includegraphics[height=.4\textwidth]{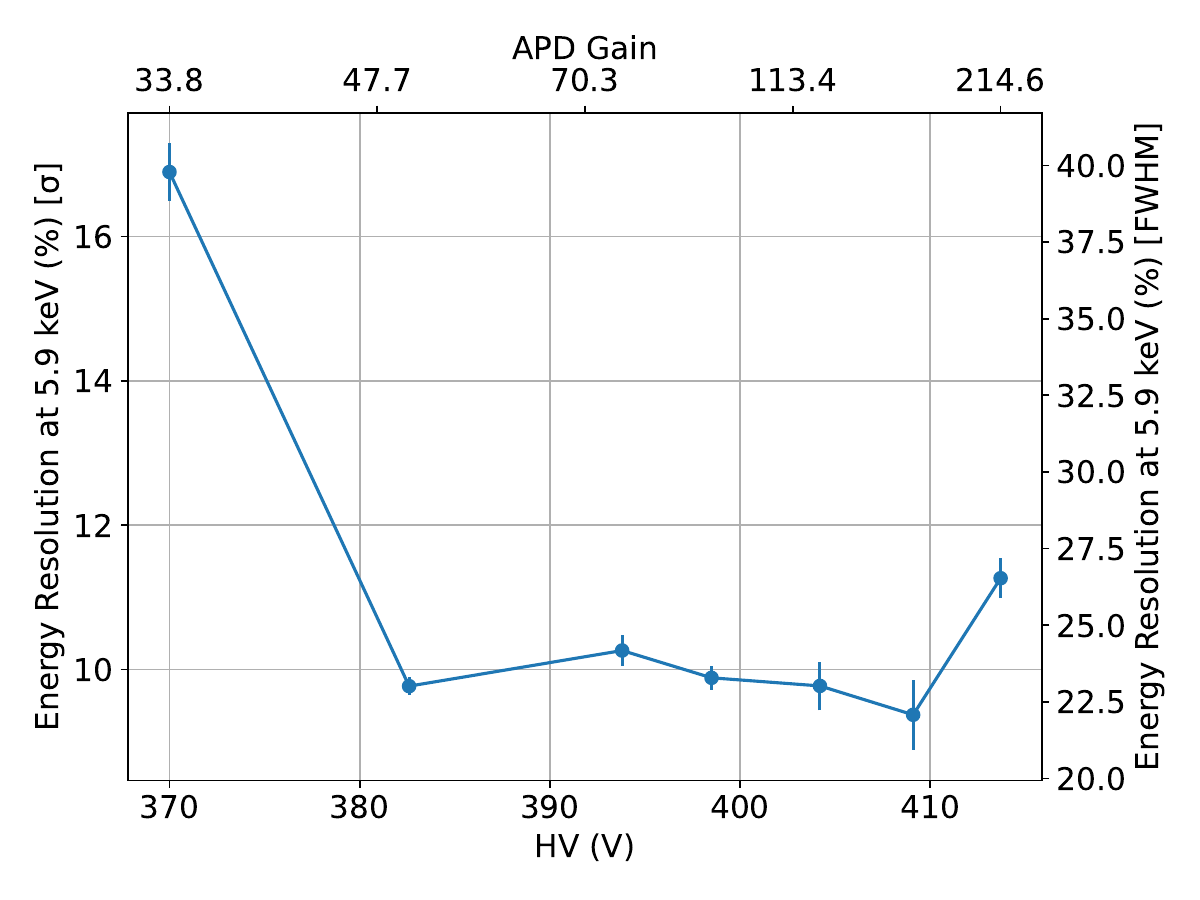}
 \caption{\textbf{Left:} $^{55}$Fe spectrum measured directly with the APD for various bias voltages \textbf{Right:} APD energy resolution as a function of its bias voltage and gain. The point at 370~V is artificially higher due to the fact that the $^{55}$Fe peak is reaching the threshold level and is therefore distorted.}
 \label{fig:Fe55_gain}
\end{figure}

The APD is first tested with no scintillator directly placing an $^{55}$Fe source on top of it, in which case the 5.95~keV from the iron source is absorbed in the depletion layer of the silicon in the APD, which directly acts as an X-ray detector. The 5.95~keV photopeak as a function of the APD gain is scanned by sweeping over the APD's bias voltage, as shown in Figure~\ref{fig:Fe55_gain}. The energy resolution of the APD itself is computed from these spectra, as also plotted in Figure~\ref{fig:Fe55_gain}. The energy resolution of the APD being much better than that of GAGG, one can safely assume that the main contribution to the energy resolution will come from the scintillator itself.

%calculation expected charge Fe55 on APD + same with Co57 -> use this for guessing the light yield with GAGG -> see calc by me and Ettore, 22.5fC/nm? ...

%-> also spectrum of Co57 directly on APD (6.4 and 14keV lines), check log book 11 nov 2024 -> or do not put at all

%-> add ADC vs gain/hv plot? with energy res plot -> or at least say that the peak position is linear with APD gain

\begin{figure}[htpb]
\centering
 \hspace*{-0.5cm}\includegraphics[height=.4\textwidth]{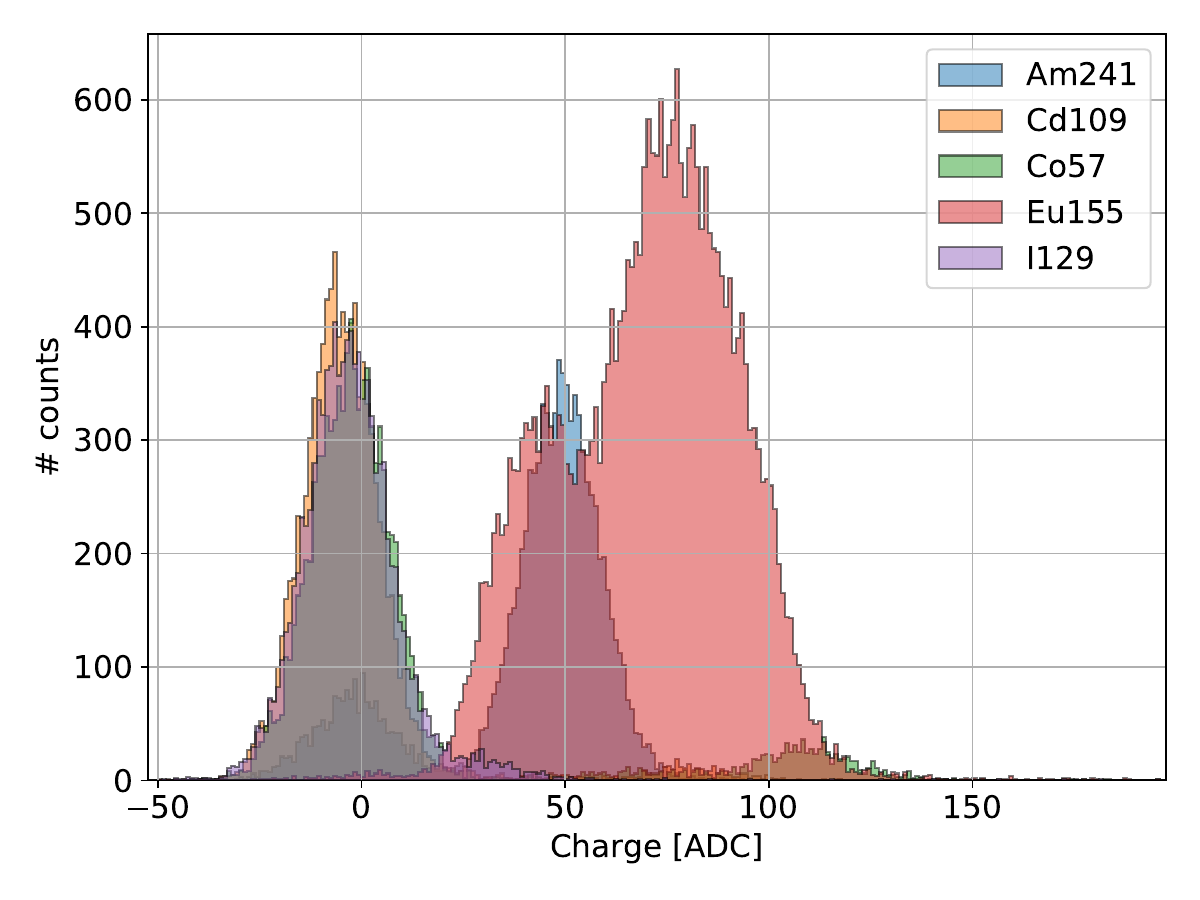}\includegraphics[height=.4\textwidth]{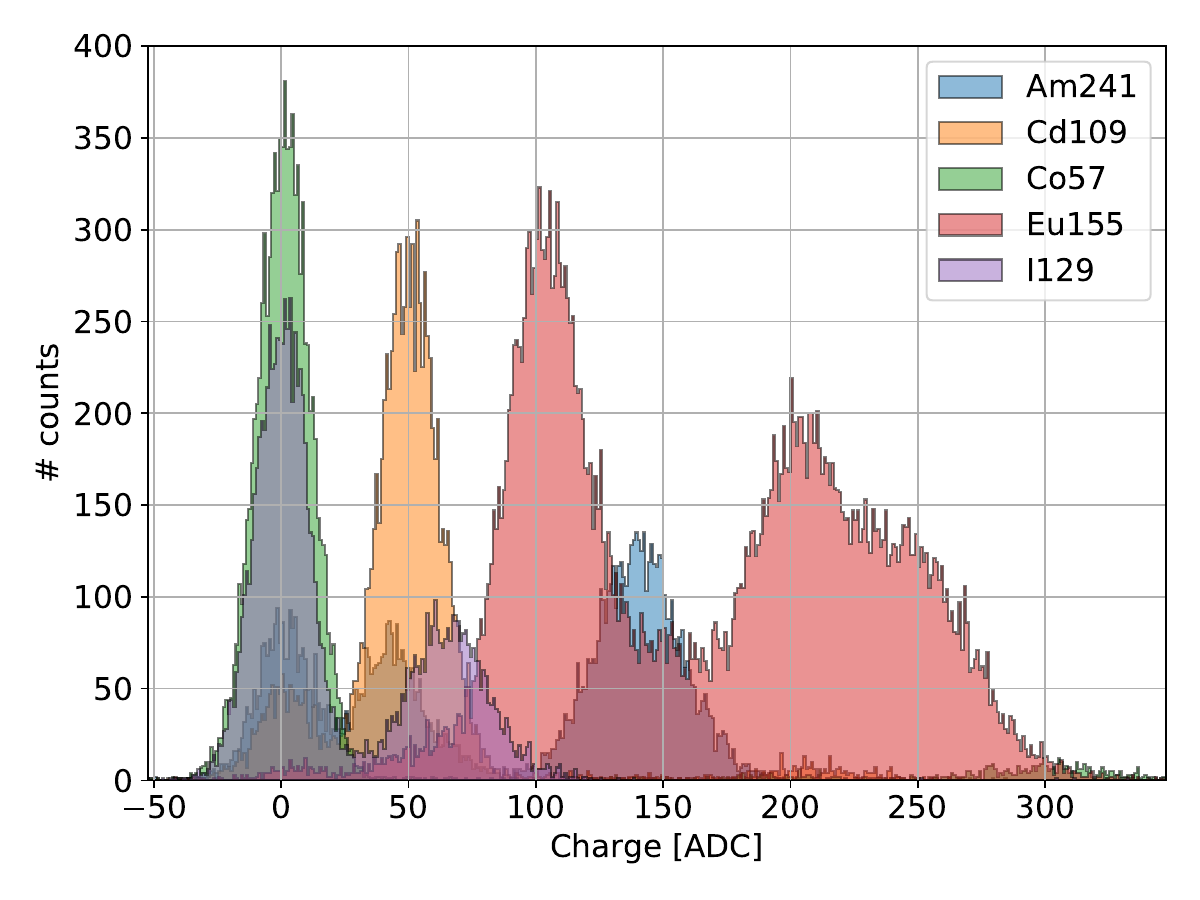}
 \caption{\textbf{Left:} Measured $^{241}$Am, $^{109}$Cd, $^{57}$Co, $^{155}$Eu, and $^{129}$I spectra for a bias voltage of 382~V. \textbf{Right:} Same spectra measured at 410~V.}
 \label{fig:all_spectra_GAGG}
\end{figure}

The absorber channel (with the APD coupled to a wrapped GAGG scintillator) is being tested using radioactive sources with characteristic emission lines in the 25-100~keV range. In addition to $^{109}$Cd that was used for the energy calibration of the scatterers, we make use of $^{241}$Am, $^{57}$Co, $^{155}$Eu, and $^{129}$I sources. Spectra acquired with the absorber for these 5 sources are shown in Figure~\ref{fig:all_spectra_GAGG} for APD bias voltages of 382 and 410~V. These spectra were subtracted for the noise pedestal of $702.41 \pm 0.14$~ADC. From the APD datasheet \citep{APDdatasheet} and dedicated APD characterization measurements \citep{APDsensors_paper}, we know that the APD gain for these two voltages is, respectively, about 50 and 100.

%The typical relation between bias voltage and gain from the APD data sheet\citep{APDdatasheet} is combined with the measured gain of 50 at 382.3~V\citep{APDsensors_paper} to the relation between gain and voltage.

\begin{figure}[htpb]
\centering
 \includegraphics[width=.85\textwidth]{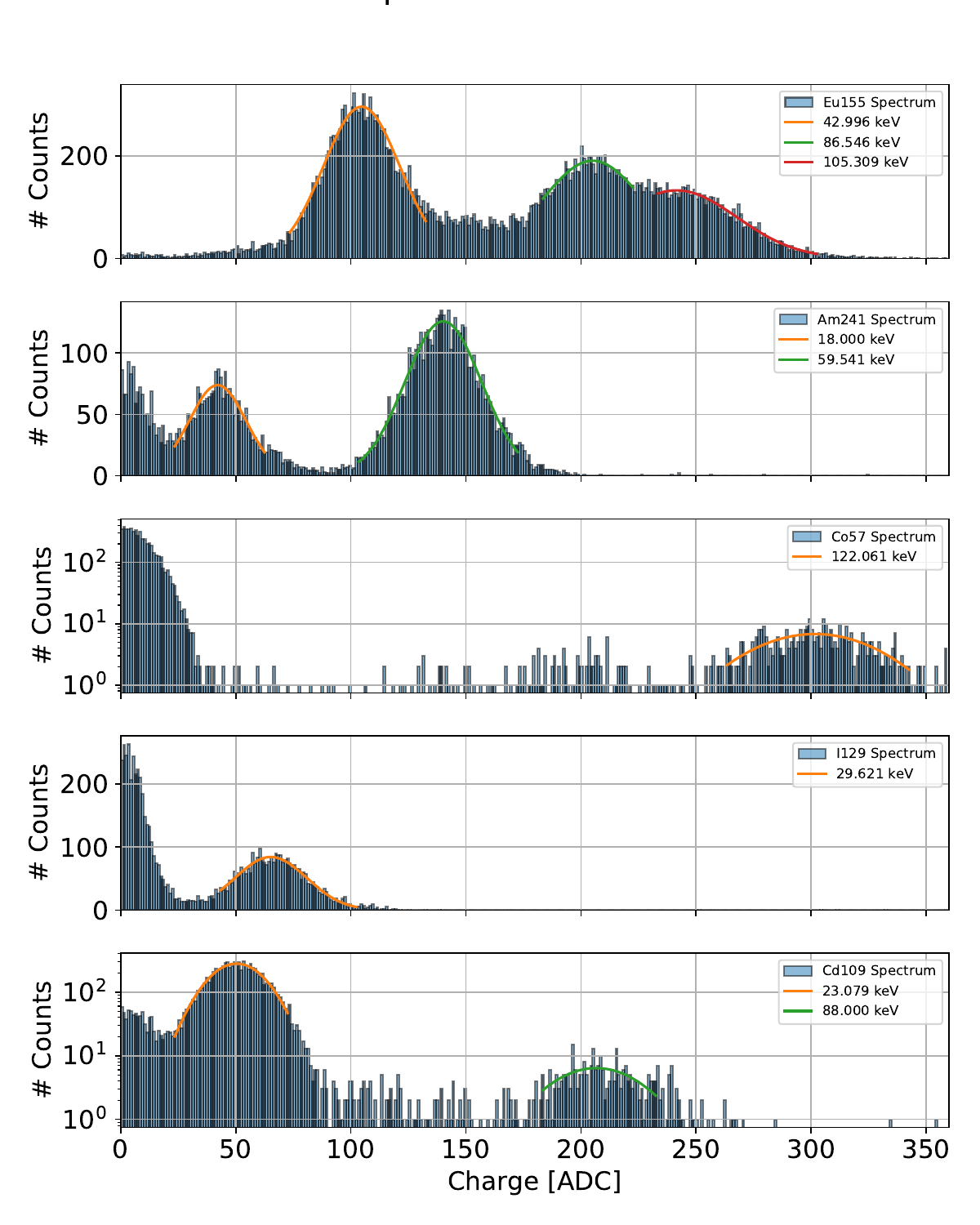}
 \caption{Spectra of the 5 sources shown in Figure \ref{fig:all_spectra_GAGG} fitted for characteristic lines, whose energies are summarized in the text.}
 \label{fig:fitted_spectra_GAGG}
\end{figure}

Fitted spectra are shown in Figure~\ref{fig:fitted_spectra_GAGG} for a bias voltage of 410~V, for which the gain is high enough to disentangle all the peaks. The $^{109}$Cd shows both the blend of lines around 23.108~keV like in the plastic as well as the 88.0~keV line. The $^{241}$Am source shows a blend of line (16.8~keV L$_{\beta2}$ + 17.74~keV L$_{\beta1}$ + 20.81~keV L$_\gamma$) with an average weighted energy of 18~keV, seen as a single line because the energy resolution does not allow for disentangling the lines. The 13.95~keV Np L$_\alpha$ is below threshold and is therefore not contributing to the measured 18~keV peak. In addition to this, the 59.541~keV line from $^{241}$Am is also seen. $^{155}$Eu has characteristic lines at 42.996, 86.546, and 105.309~keV, while a line at 29.621~keV is seen for $^{129}$I and at 122.061~keV for $^{57}$Co. The values of these characteristic energies are taken from \citep{NuDat, x-ray_booklet, abbene2023high}.

\begin{figure}[htpb]
\centering
 \hspace*{-0.5cm}\includegraphics[height=.4\textwidth]{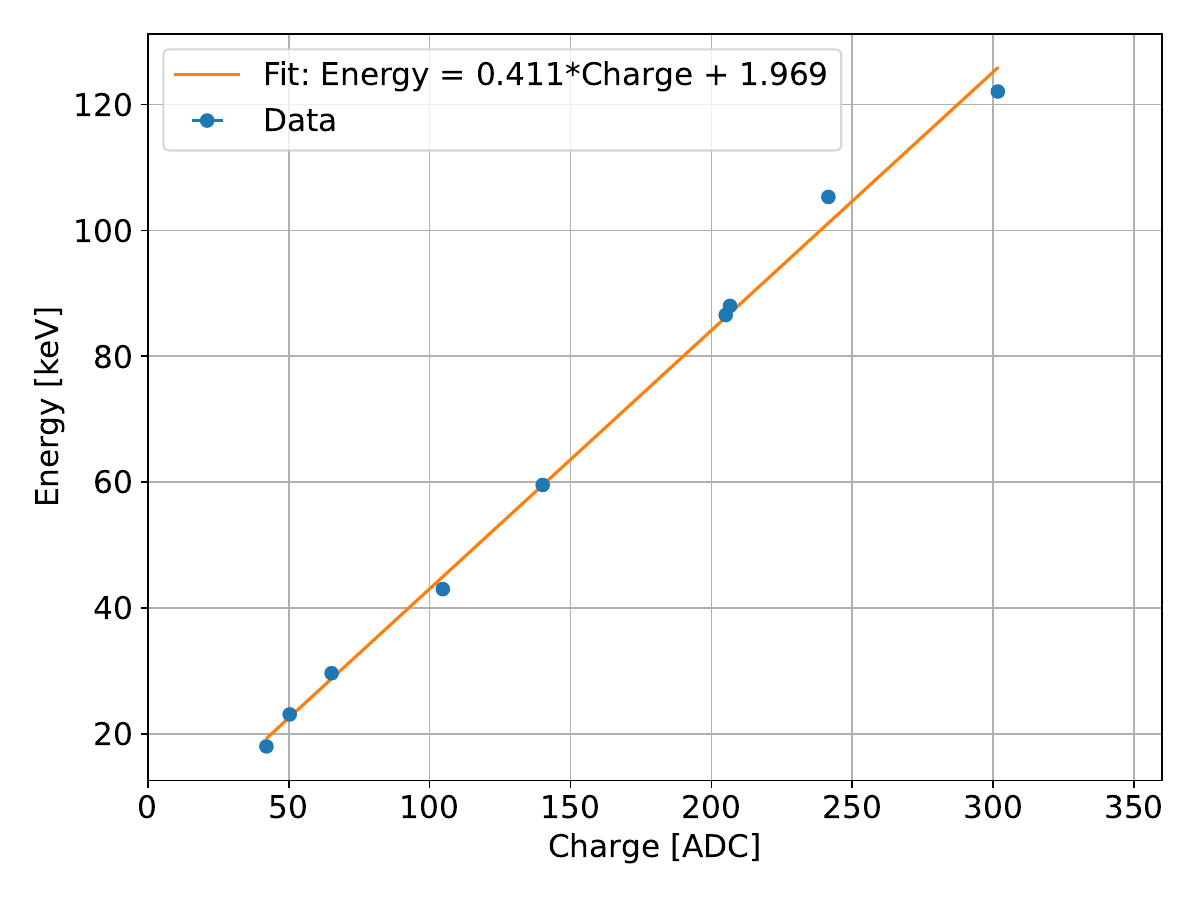}\includegraphics[height=.4\textwidth]{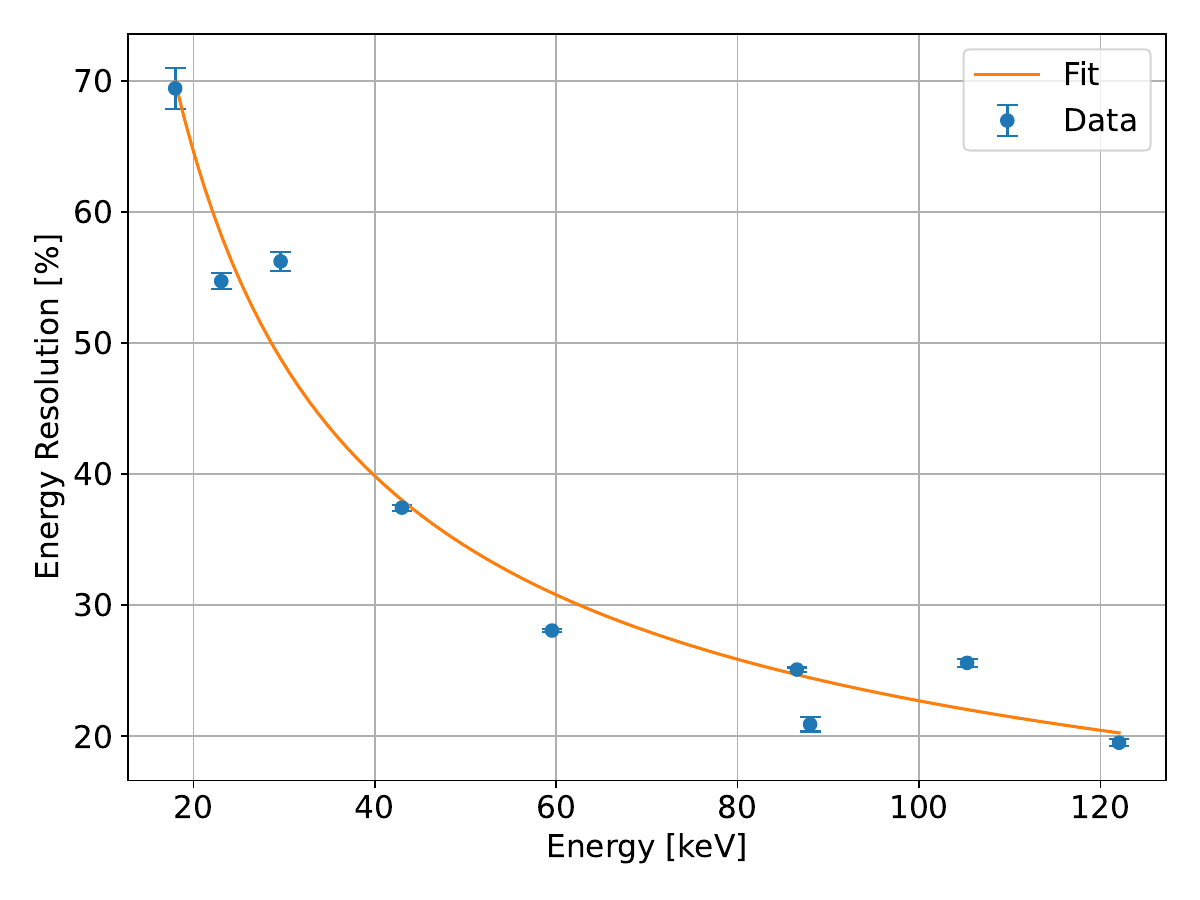}
 \caption{\textbf{Left:} Energy (keV) obtained from the theoretical emission features of the various well-known radioactive isotopes as a function of the fitted line position (ADC). \textbf{Right:} Energy resolution (FWHM) of the GAGG absorber as a function of energy.}
 \label{fig:energy_vs_charge_GAGG_energy_res}
\end{figure}

These fitted characteristic energies allow for converting the digitized charge in ADC into deposited energy in keV over the entire absorber energy range. The linearity of the charge to energy conversion over the energy band is given in Figure~\ref{fig:energy_vs_charge_GAGG_energy_res} together with the energy resolution as a function of energy computed from the full-width at half-maximum (FWHM) for each measured photopeak. The energy resolution dependency in energy is well described by the square root of a second-order polynomial in 1/E, and is comparable to the energy resolution at room temperature of GAGG crystals reported in the literature.

\begin{figure}[htpb]
\centering
 \hspace*{-0.5cm}\includegraphics[height=.4\textwidth]{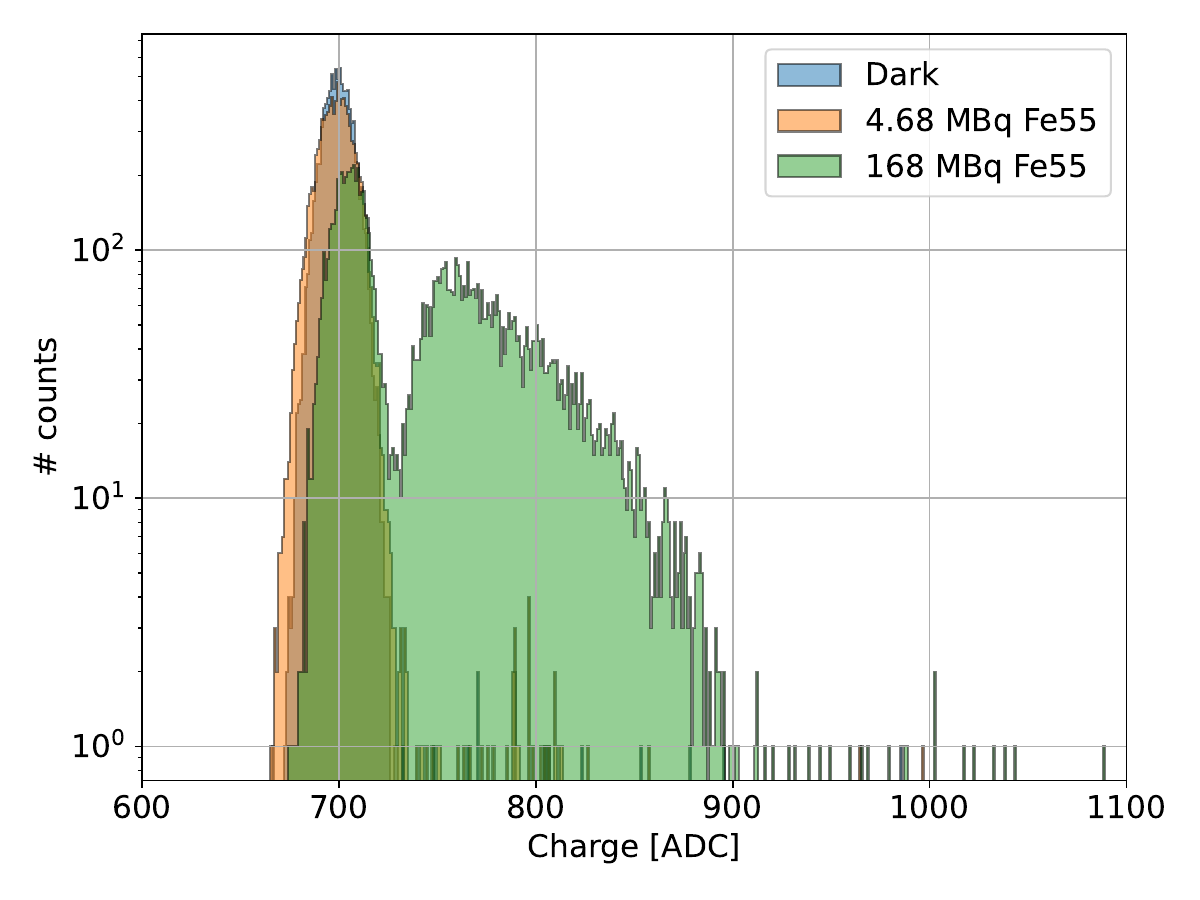}\includegraphics[height=.4\textwidth]{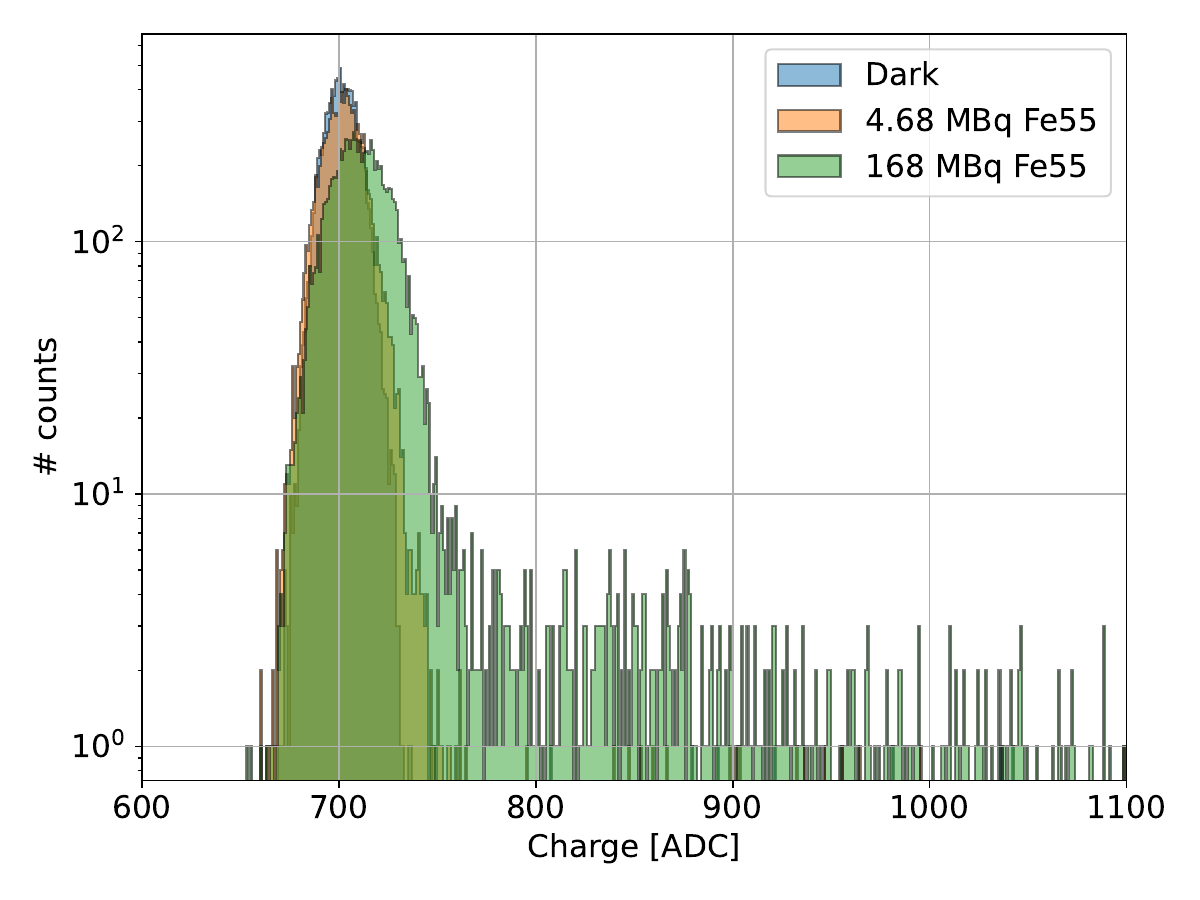}
 \caption{\textbf{Left:} $^{55}$Fe spectrum for two sources with different activities (4.68 and 168~MBq) compared to a dark spectrum at a bias voltage of 382~V. The $^{55}$Fe's 5.95~keV line is below threshold, but pile-up effects induce triggered events for a very high rate of events. \textbf{Right:} Same at a bias voltage of 410~V. The effect is less visible as the system starts to trigger on single 5.95~keV photons due to the higher gain, which dominates the spectrum.}
 \label{fig:pileup_APD}
\end{figure}

Pile-up effects due to high data rate have been studied by placing an $^{55}$Fe source on top of the GAGG crystal, for which the energy deposition in the scintillator from the 5.95~keV line is under threshold. However, as shown in Figure~\ref{fig:pileup_APD} in the spectra acquired with both 382 and 410~V bias voltages, a high data rate emulated by a high-activity $^{55}$Fe source can cause pile-up events over threshold. Indeed, a triggered peak can be observed for a voltage of 382~V using a source with an activity of 168~MBq, as several 5.95~keV energy depositions happen in the GAGG in a time scale smaller than the electronics integration time (50-100~ns, with a crystal decay time of $\sim$90~ns), which ends up in a signal over threshold due to the accumulation of scintillation light seen as a single event by the system. This effect is less prominent at 410~V as the higher gain allows the system to trigger on single 5.95~keV photons, whose rate dominates over the pile-up event rate. This pile-up effect should not affect the instrument's operation since a maximum event rate of tens of kHz is expected for bright flares. It should, however, be further studied with the next prototype design to ensure it has no significant impact on the detector's spectro-polarimetric response for bright solar flares.

\subsection{Coincidence}\label{subsec:coincidence}

\begin{figure}[htpb]
\centering
 \includegraphics[width=.8\textwidth]{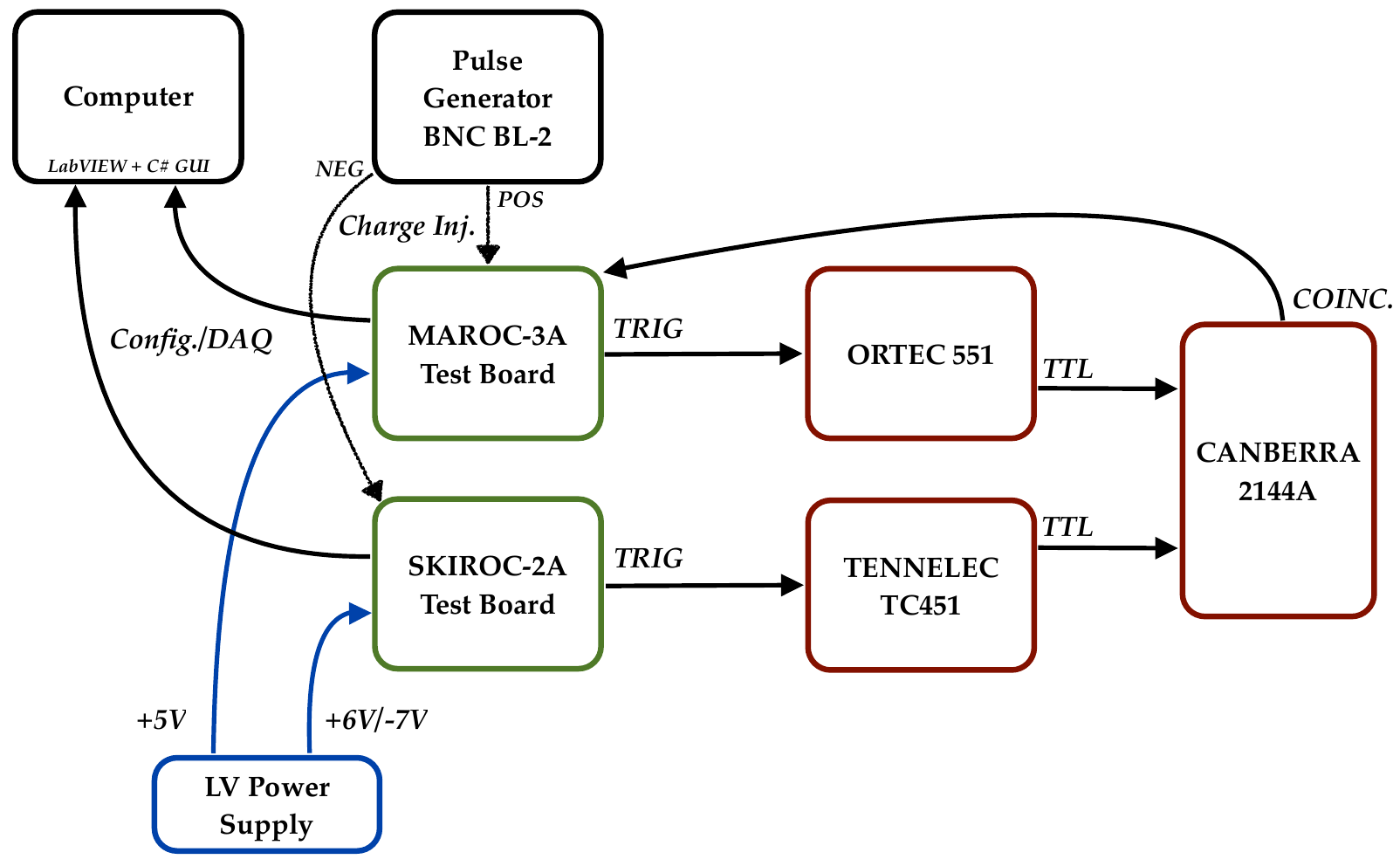}
 \caption{Diagram of the setup used for testing the Compton coincidence logic through pulse injection with both ASIC's development boards.}
 \label{fig:coinc_setup}
\end{figure}

The working principle of a Compton polarimeter implies reading in coincidence the scattering and absorption events. More specifically, polarization events, for which the azimuthal scattering direction can be determined, consist of a double event with, in the nominal topology of events\footnote{Other topologies of events, such as plastic+plastic, i.e. a photon scattering in a plastic bar and then being absorbed in another plastic bar, can be considered to increase the instrument's effective area for polarimetry. One should, however, note that double plastic events, as for a single-phase polarimeter, show a lower modulation factor and are therefore less sensitive to the polarization of the incoming photon. The possible use of these events will be assessed in future phases on the basis of measurements and simulations.}, one hit in a plastic bar and one in a GAGG scintillator. The functionality of the coincidence logic using the two ASIC development boards was tested using NIM modules for signal conversion and applying the coincidence logic when receiving an event from both parallel acquisition chains. The diagram of this setup, which showed the correct functionality of the logic, is shown in Figure~\ref{fig:coinc_setup}. The BNC BL-2 Analog pulse generator is used to inject a coincident signal in both evaluation boards. The trigger signals from both boards are converted to TTL through the ORTEC 551 and TENNELEC TC451 modules. Both TTL signals are then provided to the CANBERRA 2144A coincidence module, the output of which is fed to the MAROC-3A board as a HOLD signal to look for a scatterer signal when an absorber channel is triggered on the SKIROC-2A board.

\begin{figure}[htpb]
\centering
 \includegraphics[height=.48\textwidth]{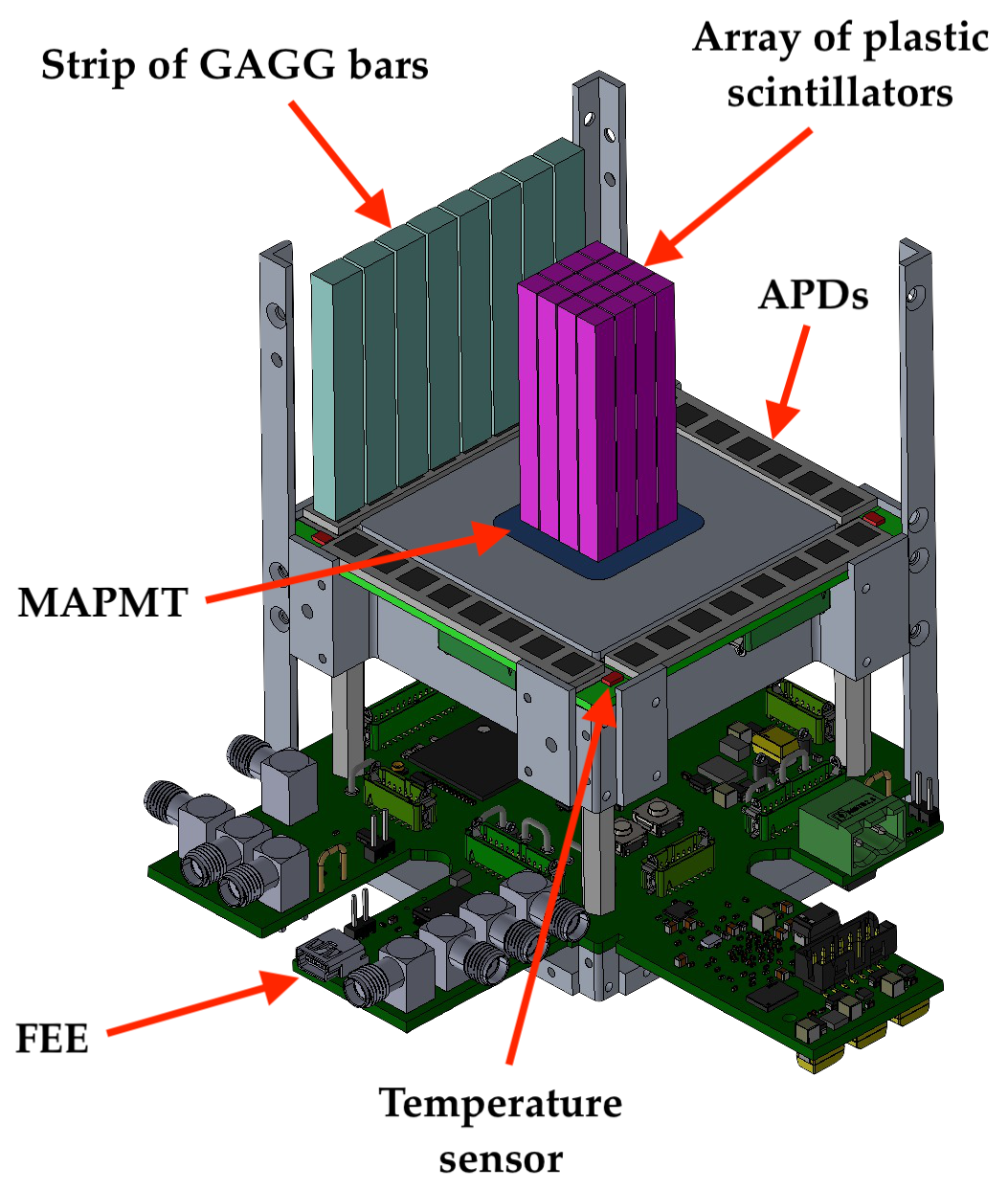}\includegraphics[height=.48\textwidth]{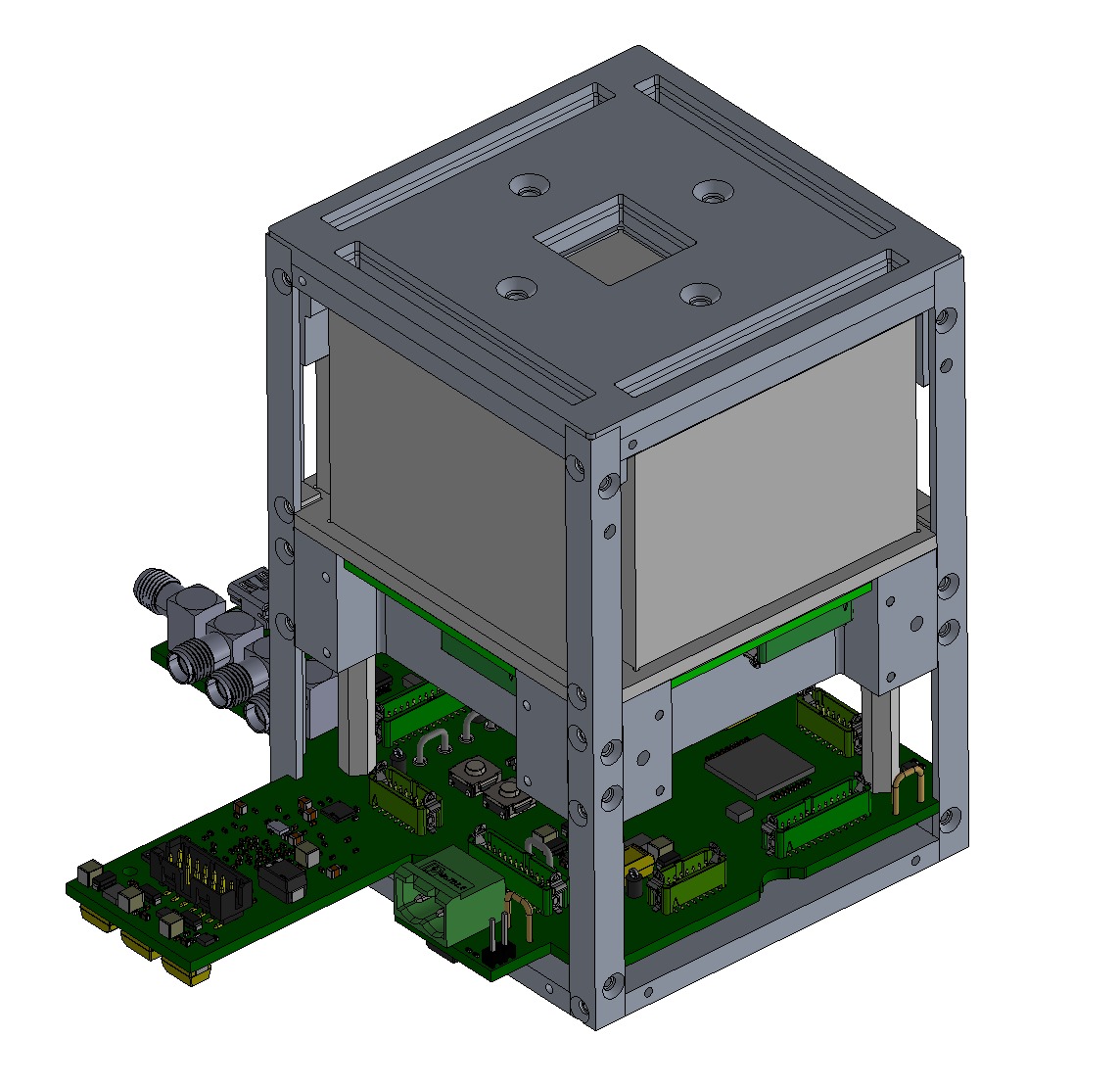}
 \caption{\textbf{Left:} CAD design showing the sensitive parts of the polarimeter prototype being developed for CUSP's phase B. \textbf{Right:} CAD design of the full assembly of the prototype.}
 \label{fig:CUSP_proto}
\end{figure}

A custom front-end electronics (FEE) based on both MAROC-3A and SKIROC-2A ASICs controlled and read out by an Artix\textsuperscript{\textregistered}-7 FPGA from Xilinx is currently under development by the Weeroc company as part of a polarimeter prototype, whose current design is shown in Figure~\ref{fig:CUSP_proto}. The prototype will contain four strips of 8 GAGG bars, like in the flight design, surrounding a single MAPMT coupled to a 4$\times$4 array of plastic scintillators. This prototype will offer better performances thanks to its dedicated electronics, and will allow us to characterize more deeply the spectral performances of the instrument as well as its spurious polarization response, or in other words, the instrumental effects distorting the polarimetric response. This prototype is currently being built as part of phase B of the CUSP mission. Both the scatterer and absorber scintillator blocks will be assembled in an epoxy-based mechanical structure hosting strips of enhanced specular reflector (ESR) films for scintillator wrapping, which offers a higher reflectivity than Teflon \citep{De_Angelis_2025_optsim}.

\section{Conclusions}

% under the Alcor program for CubeSats from the Italian Space Agency (ASI)
The CUbesat Solar Polarimeter (CUSP) is a 6U-XL mission being developed by a collaboration led by the Italian National Institute for Astrophysics of Rome (INAF-IAPS). It will host a dual-phase Compton polarimeter sensitive in the 25-100~keV with the aim of performing hard X-ray polarimetry of solar flares. In particular, CUSP will provide unprecedented constraints on the magnetic field geometry, particle acceleration, and energy release mechanisms in solar flares. With its high sensitivity, time resolution, and dedicated observing strategy, CUSP will bridge the gap between theoretical models and observational data, advancing both fundamental heliophysics and applied space weather research. Beyond solar physics, CUSP will also be able to perform spectro-polarimetric observations of bright X-ray sources that enter its field-of-view, widening its science impact to other fields of astrophysics. It will, for instance, be able to perform significant polarization measurements of bright sources such as Gamma-Ray Bursts, for which polarimetry stands as one of the key techniques for future studies to deepen our understanding of these sources \citep{Bozzo2024, DeAngelis2025Mondello}. 

Preliminary spectral performance results for both the scatterer and absorber detectors based on single-channel prototypes using ASIC development boards for data acquisition have been reported. Although the design of these single-channel prototypes was far from optimized, the results demonstrated the functionality of both acquisition chains as well as promising spectral sensitivity, with room for improvement at low energies with dedicated electronics in the future. Indeed, an energy threshold of 3.55~keV was reached using development boards, which are not optimized for performance measurements. The required 1.17~keV energy threshold for a 25~keV photon with a 90$^\circ$ angle polar scattering angle will require an improvement of the noise rejection, and should be reasonably achieved with the custom electronics in development for CUSP. A careful characterization of the detector's spectral and polarimetric responses will be performed on the prototype polarimeter to ensure that all systematic effects are understood and kept under control. No pile-up is expected as the electronics is fast-enough to acquire data from the brightest flares. The gain will be kept stable by correcting the bias voltage with the monitored temperature, and other contributions to the systematics such as optical crosstalk or residual threshold and gain non-uniformities will be accounted for in the polarimeter's response to prevent false polarization detections.

With a launch currently targeted for late 2027/early 2028, depending on ASI approval, the CUSP mission is now at the prototyping phase (phase B). While spectral performances can only be studied with single-channel detectors, a representative prototype of the flight Compton polarimeter is under construction for performing a detailed characterization of the spectro-polarimetric response of the instrument through laboratory measurements and refining our sensitivity estimates to the polarization of solar flares and other bright sources of interest.

%%%%%%%%%%%%%%%%%%%%%%%%%%%%%%%%%%%%%%%%%%
\vspace{6pt} 

%%%%%%%%%%%%%%%%%%%%%%%%%%%%%%%%%%%%%%%%%%
%% optional
%\supplementary{The following supporting information can be downloaded at:  \linksupplementary{s1}, Figure S1: title; Table S1: title; Video S1: title.}


% Only used for preprtints:
% \supplementary{The following supporting information can be downloaded at the website of this paper posted on \href{https://www.preprints.org/}{Preprints.org}.}

%\begin{tabularx}{\textwidth}{lll}
%\toprule
%\textbf{Name} & \textbf{Type} & \textbf{Description} \\
%\midrule
%S1 & Python script (.py) & Script of python source code used in XX \\
%S2 & Text (.txt) & Script of modelling code used to make Figure X \\
%S3 & Text (.txt) & Raw data from experiment X \\
%S4 & Video (.mp4) & Video demonstrating the hardware in use \\
%\bottomrule
%\end{tabularx}
%}

%%%%%%%%%%%%%%%%%%%%%%%%%%%%%%%%%%%%%%%%%%
\authorcontributions{N.D.A. performed the laboratory measurements, data analysis, and wrote this manuscript. A.K. helped with the laboratory measurements for the GAGG absorbers. S.F. and E.D.M. assisted with setting up the laboratory setup and provided guidance on laboratory measurements and data analysis. E.C. and P.S. provided advice on laboratory measurements and data analysis. G.L. and A.R. helped with setting up the laboratory setup. A.K., S.F., E.C., and R.C. helped reviewing the manuscript. All other authors are CUSP team members responsible for payload development and simulations (INAF), the satellite platform (IMT s.r.l.), the payload electronics (DEDA Connect s.r.l.), mission analysis (University of Bologna), the ground segment (University of "La Tuscia"), or project control (ASI). All authors have read and agreed to the published version of the manuscript.}

\funding{This work is funded by the Italian Space Agency (ASI) within the Alcor Program, as part of the development of the CUbesat Solar Polarimeter (CUSP) mission under ASI-INAF contract n. 2023-2-R.0.}

\dataavailability{Data is contained within the article.}

\conflictsofinterest{The activities reported in this paper are funded by ASI which also approved its publication.} 

%%%%%%%%%%%%%%%%%%%%%%%%%%%%%%%%%%%%%%%%%%
%% Optional

%% Only for journal Encyclopedia
%\entrylink{The Link to this entry published on the encyclopedia platform.}

\abbreviations{Abbreviations}{
The following abbreviations are used in this manuscript:
\\

\noindent 
\begin{tabular}{@{}ll}
APD & Avalanche PhotoDiode\\
ASI & Agenzia Spaziale Italiana (Italian Space Agency)\\
ASIC & Application Specific Integrated Circuit\\
CAD & Computer-Aided Design\\
CME & Coronal Mass Ejection\\
CUSP & CUbesat Solar Polarimeter\\
ESR & Enhanced Specular Reflector\\
FEE & Front-End Electronics\\
GAGG & Gadolinium Aluminum Gallium Garnet\\
GUI & Graphical User Interface\\
HXR & Hard X-Ray\\
MAPMT & Multi-Anode PhotoMultiplier Tube\\
MDP & Minimum Detectable Polarization\\
PMT & PhotoMultiplier Tube\\
PTFE & PolyTetraFluoroEthylene \\
PVT & PolyVinylToluene\\
SMD & Surface-Mount Device\\
UBA & Ultra Bi-Alkali
\end{tabular}
}

%%%%%%%%%%%%%%%%%%%%%%%%%%%%%%%%%%%%%%%%%%
%\isPreprints{}{% This command is only used for ``preprints''.
\begin{adjustwidth}{-\extralength}{0cm}

\reftitle{References}


\bibliography{references}

%=====================================
% References, variant B: internal bibliography
%=====================================

% If authors have biography, please use the format below
%\section*{Short Biography of Authors}
%\bio
%{\raisebox{-0.35cm}{\includegraphics[width=3.5cm,height=5.3cm,clip,keepaspectratio]{Definitions/author1.pdf}}}
%{\textbf{Firstname Lastname} Biography of first author}
%
%\bio
%{\raisebox{-0.35cm}{\includegraphics[width=3.5cm,height=5.3cm,clip,keepaspectratio]{Definitions/author2.jpg}}}
%{\textbf{Firstname Lastname} Biography of second author}

% For the MDPI journals use author-date citation, please follow the formatting guidelines on http://www.mdpi.com/authors/references
% To cite two works by the same author: \citeauthor{ref-journal-1a} (\citeyear{ref-journal-1a}, \citeyear{ref-journal-1b}). This produces: Whittaker (1967, 1975)
% To cite two works by the same author with specific pages: \citeauthor{ref-journal-3a} (\citeyear{ref-journal-3a}, p. 328; \citeyear{ref-journal-3b}, p.475). This produces: Wong (1999, p. 328; 2000, p. 475)

%%%%%%%%%%%%%%%%%%%%%%%%%%%%%%%%%%%%%%%%%%
%% for journal Sci
%\reviewreports{\\
%Reviewer 1 comments and authors’ response\\
%Reviewer 2 comments and authors’ response\\
%Reviewer 3 comments and authors’ response
%}
%%%%%%%%%%%%%%%%%%%%%%%%%%%%%%%%%%%%%%%%%%
\PublishersNote{}
%\isPreprints{}{% This command is only used for ``preprints''.
\end{adjustwidth}
\end{document}